\pdfoutput=1

\documentclass[11pt,a4paper]{article}

\usepackage{jheppub}
\usepackage{amsmath}
\usepackage{amssymb}
\usepackage{appendix}
\usepackage{multirow}
\usepackage{rotating}
\usepackage{diagbox}
\usepackage{slashed}
\usepackage{float}
\usepackage{xcolor}
\usepackage{cancel}
\usepackage{tabularx}
\usepackage{tikz}
\usepackage{bbold}
\usepackage{tikz-feynman}
\usetikzlibrary{calc,tikzmark,fit,shapes.geometric,matrix,decorations.markings,arrows.meta,decorations.pathmorphing,patterns,positioning,snakes}

\tikzset
  {midarrow/.style={decoration={markings,mark=at position 0.5 with
     {\arrow[thin,xshift=2pt]{Triangle[length=4pt,#1]}}},postaction={decorate}}
  }

\tikzset{
proton/.style = {circle, draw=black, thin, fill=black!20!white, minimum size=#1,
              inner sep=0pt, outer sep=0pt},
proton/.default = 6pt 
}

\tikzset{
blob/.style = {circle, draw=black, thin, preaction={fill, black!20!white}, pattern=north east lines, minimum size=#1,
              inner sep=0pt, outer sep=0pt},
blob/.default = 6pt 
}

\tikzset{
wc/.style = {circle, fill, minimum size=#1,
              inner sep=0pt, outer sep=0pt},
wc/.default = 4pt 
}

\tikzset{vector/.style={decorate, decoration=snake}}

\usepackage[capitalize]{cleveref}
\usepackage[normalem]{ulem}

\usepackage{color}

\newcommand{\bea}{\begin{eqnarray}}
\newcommand{\eea}{\end{eqnarray}}

\newcommand{\CV}{C_{V_L}}
\newcommand{\CS}{C_{S_L}}
\newcommand{\CSs}{C_{S_L}^{*}}
\newcommand{\CT}{C_T}
\newcommand{\CTs}{C_T^*}

\usepackage{color}
\definecolor{niceblue}{rgb}{0,0,1}
\definecolor{nicered}{rgb}{0.7,0.1,0.1}
\definecolor{nicegreen}{rgb}{0.1,0.5,.1}

\usepackage{graphicx}
\usepackage{subcaption}
\graphicspath{{figures/}{./}}

\title{Explaining the $B_{d,s}\rightarrow {K^{(*)}\bar K^{(*)}}$
 non-leptonic puzzle and charged-current $B$-anomalies via scalar leptoquarks
 }
 
\author[a]{Javier M. Lizana,}
\author[b]{Joaquim Matias,}
\author[a]{Ben A. Stefanek.}

\affiliation[a]{Physik-Institut, Universit\"at Z\"urich, CH-8057 Z\"urich, Switzerland}
\affiliation[b]{Universitat Aut\`onoma de Barcelona, 08193 Bellaterra, Barcelona,\\
Institut de F\'{i}sica d'Altes Energies (IFAE), The Barcelona Institute of Science and Technology, Campus UAB, 08193 Bellaterra (Barcelona)}

\abstract{
We present a model based on $S_1$ scalar leptoquarks to solve the tension observed in the recently proposed non-leptonic optimized observables $L_{K^{*} \bar{K}^{*}}$ and  $L_{K \bar{K}}$. These observables are constructed as ratios of U-spin related decays based on $B_{d,s}^0\rightarrow {K^{(*)0}\bar K^{(*)0}}$. 
The model gives a one-loop contribution to the Wilson coefficient of the chromomagnetic dipole operator needed to explain the tension in both non-leptonic observables, while naturally avoiding large contributions to the corresponding electromagnetic dipoles. The necessary chiral enhancement comes from an $O(1)$ Yukawa coupling with a TeV-scale right-handed neutrino running in the loop. We endow the model with a $U(2)$ flavor symmetry, necessary to protect light-family flavor observables that otherwise would be in tension. Furthermore, we show that the same $S_1$ scalar leptoquark is capable of simultaneously explaining the hints of lepton flavor universality violation observed in charged-current $B$-decays. The model therefore provides a potential link between two puzzles in $B$-physics and TeV-scale neutrino mass generation. Finally, the combined explanation of the $B$-physics puzzles unavoidably results in an enhancement of $\mathcal{B}(B\rightarrow K \nu \bar \nu)$, yielding a value close to present bounds.
}

\emailAdd{jlizana@physik.uzh.ch}
\emailAdd{matias@ifae.es}
\emailAdd{ben.stefanek@physik.uzh.ch}

\preprint{ZU-TH 25/23}
\makeatletter
\gdef\@fpheader{}
\makeatother

\begin{document}

\maketitle
\section{Introduction and Motivation}
\label{intro}

Even if the hints of New Physics (NP) in  neutral ($b\to s \ell\ell$)
Lepton flavor  Universality Violating (LFUV) observables are substantially weaker than a year ago due to the particle identification problems found at LHCb in the electronic channels \cite{LHCb:2022zom,
LHCb:2022qnv}, the corresponding LFUV observables of charged ($b\to c \ell\nu$) decays \cite{HeavyFlavorAveragingGroup:2022wzx,LHCb:2023zxo},  the systematic deficit in $b\to s \mu\mu$ branching ratios~\cite{LHCb:2014cxe,LHCb:2016ykl,LHCb:2021zwz} and some of their optimized angular observables like $P_{5\mu}^\prime$~\cite{Descotes-Genon:2012isb,
LHCb:2020lmf,LHCb:2020gog} 
remain to be explained. Indeed it is interesting to emphasize that these remaining anomalies (charged-current and $b \to s\mu\mu$) can be consistently connected in a particular NP scenario at the EFT level assuming a left-handed NP explanation of the charged-current anomaly, also predicting a large enhancement to $b\to s\tau\tau$~\cite{Capdevila:2017iqn}. This scenario was proposed in \cite{Alguero:2018nvb} and leads to a LFU contribution via renormalization group evolution (RGE) to the $b\rightarrow s\ell\ell$ semi-leptonic operator coupling vectorially to leptons. 
Moreover, as it was pointed out in~\cite{Alguero:2023jeh}, this vectorial LFU contribution to $b\rightarrow s\ell\ell$ transitions can naturally accommodate the new experimental situation with rather SM-like values for the $B_s \to \mu^+ \mu^-$ average and the LFUV observables $R_{K,K^*}$~\cite{Hiller:2003js}.

It is well known that LFUV observables are a test of the universality of the coupling of  gauge bosons to electrons, muons and tau leptons. In particular, the observable  $R_K$ compares the semi-leptonic branching ratio $B^+ \to K^+\ell\ell$ between 2nd generation  ($\ell=\mu$) and 1st generation  ($\ell=e$) leptons.
Following this idea, we consider it is worthwhile to keep searching for NP in a different type of observable that also compares second versus first 
generation, but with the $d$-quark playing the role of electrons and the $s$-quark the one of muons. In this case, the relation between the two decays is driven not by lepton flavor universality, but rather by U-spin.
This is the realm of non-leptonic $B$ decays, a more theoretically complicated region than the much simpler semi-leptonic $B$ decays.

In this context it was pointed out in \cite{Biswas:2023pyw,
     Alguero:2020xca} that one can construct two non-leptonic observables, 
called $L_{K^*\bar{K}^*}$ and $L_{K\bar{K}}$,
 using the ratio of longitudinal branching ratios of the $\bar{B}_s^0  \to K^{*0}\bar{K}^{*0}$ decay versus the corresponding $\bar{B}_d^0 \to K^{*0}\bar{K}^{*0}$ decay and, similarly, using the ratio of branching ratios of the decays $\bar{B}_{d,s}^0 \to K^0\bar{K}^0$ . These so-called $L$-observables exhibit a tension with respect to their SM prediction of 2.6$\sigma$ and 2.4$\sigma$, respectively.

In this work, we present a model based on scalar leptoquarks that can simultaneously explain the deviations observed in these non-leptonic observables as well as in charged-current $B$-decays (measured via the LFU ratios $R_{D^{(*)}}$), while remaining in agreement with all relevant constraints. Therefore, in this case we explore a possible link between the charged-current anomalies and non-leptonic ones, while leaving aside the neutral-current $b \to s \ell\ell$ anomalies. We also comment on alternative, but more contrived solutions to the non-leptonic puzzle and summarize the main problems with these solutions. See also \cite{Li:2022mtc} for another proposed model-building solution to the observed non-leptonic tension based on a non-universal $Z^\prime$.

This paper is structured as follows: In \cref{sec:obsNPEFT} we review the main structure of the non-leptonic observables and their NP sensitivity. In \cref{sec:coloron} we briefly describe a coloron model and its problems while in \cref{sec:leptoquark} we present our scalar leptoquark (LQ) model for the non-leptonic observables, focusing on its impact in the relevant Wilson coefficients (WCs).  In \cref{sec:connect2CC} we discuss all relevant constraints and focus on links to other interesting observables, particularly $R_{D^{(*)}}$. Finally, we present our conclusions in \cref{sec:conc}. Some more detailed information relevant for the discussion of constraints is provided in the appendices.


\section{Brief description of the observables and EFT New Physics sensitivity}
\label{sec:obsNPEFT}
Non-leptonic $B$ decays offer another handle to test the presence of NP in flavor physics. However, these type of decays (contrary to decays to two leptons where hadronic uncertainties are minimized, but similarly to semi-leptonic $B$ decays) require optimization of the observables to reduce their hadronic sensitivity and maximize their NP sensitivity. As in the semi-leptonic case, we use a weak effective field theory description to quantify NP sensitivity. In particular, the effective Hamiltonian to describe $b\to s$ transitions in non-leptonic $B$-decays is given by
\begin{equation}\label{eq:wet}
H_{\rm eff}=\frac{G_F}{\sqrt{2}}\sum_{p=c,u} \lambda_p^{(q)}
 \Big({\cal C}_{1s}^{p} Q_{1s}^p + {\cal C}_{2s}^{p} Q_{2s}^p+\sum_{i=3 \ldots 10} {\cal C}_{is} Q_{is} + {\cal C}_{7\gamma s} Q_{7\gamma s} + {\cal C}_{8gs} Q_{8gs}\Big) \,, 
\end{equation}
where the only operators relevant for the present discussion are:
\begin{align}
 Q_{4s} &= (\bar s_a b_b)_{V-A} \sum_q\,(\bar q_b q_a)_{V-A} \,, &
 \nonumber\\[-2.2mm]
 Q_{5s} &= (\bar s b)_{V-A} \sum_q\,(\bar q q)_{V+A} \,, &Q_{7\gamma s} &= \frac{-e}{8\pi^2}\,m_b\bar s\sigma_{\mu\nu}(1+\gamma_5) F^{\mu\nu} b \,,\nonumber \\[-2.2mm]
 Q_{6s} &= (\bar s_a b_b)_{V-A} \sum_q\,(\bar q_b q_a)_{V+A} \, , &Q_{8gs} &= \frac{-g_s}{8\pi^2}\,m_b\, \bar s\sigma_{\mu\nu}(1+\gamma_5) G^{\mu\nu} b \,, \nonumber
\end{align}
with $\lambda_p^{(q)}=V_{pb}V^*_{pq}$, $(\bar q_1 q_2)_{V\pm A}=\bar q_1\gamma_\mu(1\pm\gamma_5)q_2$,
$a,b$ are colour indices, and a summation over $q=u,d,s,c,b$ is implied. The operators $Q_{4s\ldots 6s}$ are known as QCD penguin operators, while $Q_{7\gamma s}$ and $Q_{8gs}$ are electromagnetic and chromomagnetic dipole operators, respectively. One can write a similar effective Hamiltonian for $b\to d$ transitions just changing $s$ by $d$. See Refs.~\cite{
Beneke:2001ev,Biswas:2023pyw} for the definitions and conventions of the complete basis of operators.

 Concerning the optimization of the non-leptonic observables, two observables based upon ratios of the decay modes $\bar{B}_{d,s}^0 \to K^{*0}\bar{K}^{*0}$ (longitudinal component) and $\bar{B}_{d,s}^0 \to K^0 \bar{K}^0$ were proposed and analyzed in 
\cite{Biswas:2023pyw}
and
\cite{Alguero:2020xca}. In particular, they are defined as:
\begin{eqnarray}  L_{K^*\bar{K}^*}&=&\rho(m_{K^{*0}},m_{K^{*0}})\frac{{\rm BR}_{\rm long}(\bar{B}_s^0 \to K^{*0} \bar{K}^{*0})}{{\rm BR}_{\rm long}(\bar{B}_d^0 \to K^{*0} \bar{K}^{*0})}  \,, \\  
L_{K\bar{K}}&=&\rho(m_{K^0},m_{K^0})\frac{{\rm BR}(\bar{B}_s^0 \to K^0\bar{K}^0)}{{\rm BR}(\bar{B}_d \to K^0\bar{K}^0)} \,,
\end{eqnarray}
where the function $\rho$ stands for the ratio of phase space factors (see the definition in \cite{Biswas:2023pyw}) and it is very close to one in both cases. 
These observables are constructed to reduce the sensitivity to dangerous endpoint infrared divergences coming from hard scattering and annihilation diagrams. The ratio of these decays are governed by U-spin, which is a broken symmetry in the SM. 

The computation of these observables in the SM including NLO $\alpha_s$-corrections as well as power enhanced contributions in the framework of QCD-Factorization (QCDF)~\cite{Beneke:2003zv,
Beneke:2006hg} leads to the following predictions in the SM \cite{Biswas:2023pyw}:
\begin{equation} \label{LSM}
L_{K^*\bar{K}^*}^{\rm SM}=19.53^{+9.14}_{-6.64}\,, \quad \quad L_{K\bar{K}}^{\rm SM}=26.00^{+3.88}_{-3.59} \,,
\end{equation}
while their experimental measurements read
\cite{
ParticleDataGroup:2022pth, BaBar:2007wwj,
LHCb:2019bnl,
BaBar:2006enb,
Belle:2012dmz,Belle:2015gho,LHCb:2020wrt}:
\begin{equation} \label{Lexp}
L_{K^*\bar{K}^*}^{\rm exp}=4.43 \pm 0.92 \,, \quad \quad L_{K\bar{K}}^{\rm exp}=14.58\pm 3.37 \,.
\end{equation}
In both cases, a significant deficit with respect to the SM prediction is observed.
However, the relevant feature here is that both ratios (i.e. decays to vectors or pseudoscalars) can be coherently explained with the same NP contributions to the Wilson coefficients of the following two operators:
\begin{itemize}
    \item QCD-penguin operator ${\cal O}_{4q}$. Here we can consider NP only in the Wilson coefficient of the $b\to s$ transition (${\cal C}_{4s}$), or NP in both $b \to s$ and $b \to d$  (${\cal C}_{4d}$)       
    transitions. Let us now illustrate the size of the required NP contribution: in the former case a value of ${\cal C}_{4s}^{\rm NP}\approx 0.016$ is called for, while in the latter case ${\cal C}_{4s}^{\rm NP}\approx 0.010$ and ${\cal C}_{4d}^{\rm NP} \approx -0.006$ is also possible according to \cite{Biswas:2023pyw}. These values should be compared with the SM value of ${\cal C}_{4s}^{\rm SM}(\mu=4.2 {\rm GeV})=-0.036$.    
    
    \item Chromomagnetic dipole operator ${\cal O}_{8gq}$. Also here we consider two cases: NP only in $b \to s$ (${\cal C}_{8gs}^{\rm NP}$) or in both $b \to s$ 
    and $b \to d$ (${\cal C}_{8gd}^{\rm NP}$). In the former case a
 large NP contribution is required of order ${\cal C}_{8gs}^{\rm NP} \approx -0.32$, while in the latter case one can have 
 ${\cal C}_{8gs}^{\rm NP} \approx -0.16$ and ${\cal C}_{8gd}^{\rm NP} \approx +0.16$, to be compared with ${\cal C}_{8gs}^{\rm SM}(\mu=4.2 {\rm GeV})=-0.15$. Taken at face value, the present tension observed in the pure branching ratios of both $b\to s$ and $b\to d$ transitions has a preference for NP in both WCs. Finally, the relatively large NP contribution required is not a problem given the loose bounds on this WC which currently admits NP contributions as large as a few times its SM value. This is mainly because the current experimental bound is ${\rm BR}(b \to s g) < 6.8\%$~\cite{CLEO:1997xir, HeavyFlavorAveragingGroup:2022wzx} while the SM contribution is at the level of $0.5\%$~\cite{Greub:2000sy}.

\end{itemize}

Finally, we note that there is a third possibility in the pure pseudoscalar-pseudoscalar case, which is the QCD-penguin operator ${\cal O}_{6q}$. 
Here, the chirally-enhanced contribution entering these decays includes a contribution from ${\cal C}_{6s}+{\cal C}_{5s}/N_c$ that is absent in the vector-vector case, allowing for an explanation of the tension exclusively in the $L_{K\bar{K}}$ observable via contributions of this type. However, even if a non-zero ${\cal C}_{6s}$ can accommodate the tension in $L_{K\bar{K}}$, we will not consider this possibility here where we focus on a combined explanation of both $L$-observables.

In \cref{sec:coloron,sec:leptoquark} we organize the discussion 
according to the main Wilson coefficients generated by the NP model.

\section{Coloron model for the QCD penguin operator $O_{4s}$ }
\label{sec:coloron}

A previous attempt to explain the puzzle of the non-leptonic anomalies using the QCD-penguin operator ${\cal O}_{4s}$  was presented in \cite{Alguero:2020xca}.
There the contribution to ${\cal O}_{4s}$ from a 
color-octet vector (coloron) with universal flavor-diagonal couplings to the first two generations of quarks (to avoid large effects in $K$ and $D$ mixing) was explored. 
Such a particle appears naturally in extra-dimensional models as a KK-gluon, composite sectors as a resonance, or theories that extend $SU(3)_c$ to larger gauge symmetries~\cite{Chivukula:2013kw,DiLuzio:2017vat,Greljo:2018tuh}.
We consider a simplified model where the coloron couples to down quarks of different flavors according to:
\begin{equation}
{\cal L}=\Delta_{sb}^L (\bar{s} \gamma^{\mu} P_L T^a b) G_{\mu}^{\prime a} + \Delta_{sb}^R (\bar{s} \gamma^{\mu} P_R T^a b) G_{\mu}^{\prime a} \,,
\end{equation}
where similarly the flavor-diagonal couplings are denoted by $\Delta_{qq}^{L,R}$.
The contribution of such a particle to the relevant QCD-penguin Wilson coefficient\footnote{To avoid over-complicating the notation for the rest of the paper whenever a Wilson coefficient appears it should be understood as its NP contribution, namely, ${\cal C}_i^{\rm NP}$. Only its SM value will be explicitly labeled with ``SM", unless there may otherwise be confusion.
} is \cite{Alguero:2020xca}
\begin{equation} \label{WCQCD4s}
    {\cal C}_{4s}=-\frac{1}{4}\frac{\Delta_{sb}^L\Delta_{qq}^L }{\sqrt{2} G_F V_{tb} V_{ts}^* M_{G'}^2 },
\end{equation}
where $M_{G'}$ is the mass of the massive $SU(3)_c$ octet vector particle. The issue here is that di-jet searches strongly constrain the coupling of the coloron to light-generation quarks, giving the bound~\cite{CMS:2017caz}
\begin{equation}
\frac{\Delta_{qq}^L}{M_{G'}}<\frac{2.2}{10 \, {\rm TeV}}.
\end{equation}
As a consequence, in order to explain the non-leptonic anomalies the coupling $\Delta_{sb}^L$ has to be large
\begin{equation}
\frac{\Delta_{sb}^L}{M_{G'}} \approx \frac{1}{5\,{\rm TeV}},
\end{equation}
as can be seen from~\cref{WCQCD4s}.
However, such a large flavor-violating coupling is in strong tension with $\Delta M_{B_s}$: if $\Delta_{sb}^R$ is not activated, $B_s$-mixing requires~\cite{Alguero:2020xca,UTfit:2007eik,FlavConstraints}
\begin{equation}
\frac{\Delta_{sb}^L}{M_{G'}} \lesssim \frac{1}{100\,{\rm TeV}}~~~~(95\%\,{\rm CL}).
\end{equation}
The problem of the tension between di-jet searches and the $\Delta M_{B_s}$ constraint can be  partially alleviated  by allowing a substantial fine-tuning between the $\Delta_{sb}^L$ and $\Delta_{sb}^R$ contributions to $\Delta M_{B_s}$ (see \cite{Alguero:2020xca} for further details).
However, even allowing such a tuning, we find  difficult to realize such a large flavor-violating coupling (order $O(1)$ for a $5\,$TeV mass) in a natural UV-complete model, as it would need to come from $O(1)$ mixing angles between the $b$ and $s$ quarks, in constrast to the CKM-like expectation of $O(V_{ts})$.

\section{ $S_1$ scalar leptoquark solution of the non-leptonic puzzle via $O_{8gs(d)}$} \label{sec:leptoquark}
Due to the difficulty avoiding the bound on $B_s$-mixing (without a large amount of fine-tuning) in the case of the QCD penguin, we turn now to a solution of the non-leptonic puzzle via the chromomagnetic dipole operators $O_{8gs}$ and ${O}_{8gd}$. We have seen in \cref{sec:obsNPEFT} that explaining the non-leptonic puzzle via these operators requires a NP contribution that is of the same size as the SM value if there is NP in both $b \to s$ and $b\to d$ (or double if only $b \to s$ is switched on).
While dipole operators $\bar d_L^i \sigma_{\mu\nu} b_R$ always come with an $m_b$ suppression in the SM, this chiral suppression can in general be lifted in NP models. In particular, if we assume that the NP sector follows the same structure as the SM up to a (maximal) chiral enhancement of $m_t/m_b$, one is pointed to a low effective NP scale
\begin{equation}
\frac{m_t}{\Lambda_{\rm NP}^2} \approx \frac{m_b}{m_W^2} \qquad \implies \qquad \Lambda_{\rm NP} \approx m_W \sqrt{\frac{m_t}{m_b}} \approx 500~{\rm GeV} \,.
\label{eq:NPscale}
\end{equation}
We turn now to the NP origin of this chiral enhancement. If we restrict our attention to weakly-coupled beyond the SM particles with spin $\leq 1$, such dipoles must be loop-generated. In particular, since both color and the down-quark charge flow through the loop, one unavoidably also generates the corresponding electromagnetic dipole ${\cal C}_{7\gamma}$, giving rise to the $b\rightarrow s(d) \gamma$ flavor-changing neutral current (FCNC). Furthermore, the chromomagnetic and electromagnetic dipoles mix into each other under RGE. However, as we will see later, if we have ${\cal C}_{7\gamma}/{\cal C}_{8g}\approx -1/3$, one obtains an accidental cancellation in the RGE that allows for ${\cal C}_{8gs}^{\rm NP} \approx {\cal C}_{8gs}^{\rm SM}$ while passing the bound on $b\rightarrow s \gamma$. Even if this may seem like a tuning, it is achieved naturally if the electric charge flowing inside the loop follows the color flow, predicting ${\cal C}_{7\gamma}/{\cal C}_{8g} =  -1/3$ exactly before RGE effects. Furthermore, the dipole loop should consist of a fermion line as well as a bosonic one. Assuming at most 2 NP states, if color flows along the fermion line one can see that the bosonic line must be a spin-1 singlet, aka a $Z' \sim ({\bf 1,  1}, 0)$ vector. Similarly, if color flows along the bosonic line, then the boson must be a scalar carrying color and the charge of the down quark. Therefore, its SM quantum numbers should be $S_1 \sim ({\bf \bar 3,  1}, 1/3)$, a scalar leptoquark. Because this particular leptoquark can also be responsible for hints of LFUV observed in charged-current $B$-decays, we focus on this option in what follows. 

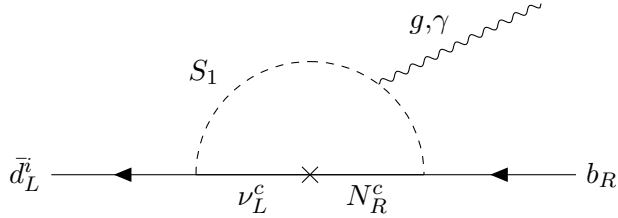
\begin{figure}[t]
    \centering
    \begin{tabular}{c@{\hskip 0.6cm}c}
\begin{tikzpicture}
\begin{feynman}
\vertex (a1) {$\bar{d}_L^i$};
\vertex[right=2.25cm of a1] (a2);
\vertex[right=3cm of a2] (a3);
\vertex[right=2cm of a3] (a4) {$b_R$};
\vertex at ($(a2)!0.5!(a3)!1.5cm!90:(a3)$) (d);
\vertex at ($(a2)!0.8!(a3)!1.2cm!90:(a3)$) (e);
\vertex at ($(a2)!0.5!(a3)!0.0cm!90:(a3)$) (f);
\vertex[above=of a4] (c1) ;
\vertex[above=2em of c1] (c3);
\vertex at ($(c1)!1!(c3) - (0.8cm, 0)$) (c2);
\diagram* {
(a4) -- [fermion] (a3) -- [solid,edge label=$N^c_R$] (f)  --  [solid,edge label=$\nu_L^c$] (a2) -- [fermion] (a1),
(a3) -- [insertion=0.5] (a2), 
(a2) -- [scalar, quarter left, edge label=$S_1$] (d) -- [scalar, quarter left] (a3),
(e) -- [boson, edge label=$g\text{,}\gamma$] (c2),
};
\end{feynman}
\end{tikzpicture}
    \end{tabular}
    \caption{Feynman diagrams contributing to $C_{7\gamma i}$ and $C_{8 g i}$, where $i=d,s$. The cross on the fermion line corresponds to an insertion of the $O(m_t)$ neutrino Dirac mass.}
     \label{fig:FeyDiag}
\end{figure}

Taking $S_1$ as our mediator, we see that the fermion line should be both color and electrically neutral. Therefore, the origin of the chiral enhancement in this scenario is an $O(1)$ Yukawa coupling between the SM lepton doublet and a right-handed (RH) neutrino $N_R$, namely $\mathcal{L} \supset - y_N \bar \ell_L \tilde H N_R$, as shown in~\cref{fig:FeyDiag} in the electroweak (EW) broken phase, where $y_N \langle H \rangle = O(m_t)$. In order to have a SM-sized effect in the chromomagnetic dipoles, both NP states ($S_1$ and $N_R$) should have masses of order the TeV scale.
\begin{figure}[h!]
    \centering
    \begin{tabular}{c@{\hskip 0.6cm}c}
         \begin{tikzpicture}[thick,>=stealth,scale=1.2,baseline=-0.5ex]
            \draw[midarrow] (-1.6,0.8) node[left] {$q_2$} -- (0,0.8)  ;
            \draw[midarrow] (0,0.8) -- (1.6,0.8) node[right] {$\ell$};
            \draw[dashed] (0,-0.8) -- (0,0.8); 
            \draw[midarrow] (0,-0.8) -- (-1.6,-0.8) node[left] {$q_1$};
            \draw[midarrow] (1.6,-0.8) node[right] {$\ell$} -- (0,-0.8);
            \node[right] at (0.01,0) {$S_1$};
        \end{tikzpicture}
        \hspace{20mm}
        \begin{tikzpicture}[thick,>=stealth,scale=1.2,baseline=-0.5ex]
            \draw[midarrow] (-1.6,0.8) node[left] {$q_{2}$} -- (-0.8,0.8);
            \draw[midarrow] (-0.8,0.8) -- (0.8,0.8);
            \draw[dashed] (0.8,0.8) -- (0.8,-0.8);
            \draw[midarrow] (0.8,-0.8) -- (-0.8,-0.8);
            \draw[midarrow] (-0.8,-0.8) -- (-1.6,-0.8) node[left] {$q_{1}$};
            \draw[dashed] (-0.8,0.8) -- (-0.8,-0.8);
            \draw[midarrow] (0.8,0.8) -- (1.6,0.8) node[right] {$q_{1}$};
            \draw[midarrow] (1.6,-0.8) node[right] {$q_{2}$} -- (0.8,-0.8);
            \node[above] at (0,0.8) {$\ell$};
            \node[below] at (0,-0.8) {$\ell$};
            \node[right] at (0.8,0) {$S_1$};
            \node[left] at (-0.8,0) {$S_1$};
        \end{tikzpicture}
    \end{tabular}
    \caption{Problematic FCNC diagrams induced when $S_1$ couples to both $q_{1}$ and $q_{2}$. Left: Tree-level contribution to $K\rightarrow \pi \nu\bar\nu$. Right: One-loop contribution to $K$-$\bar K$/$D$-$\bar D$ mixing.}
    \label{fig:S1FCNC}
\end{figure}
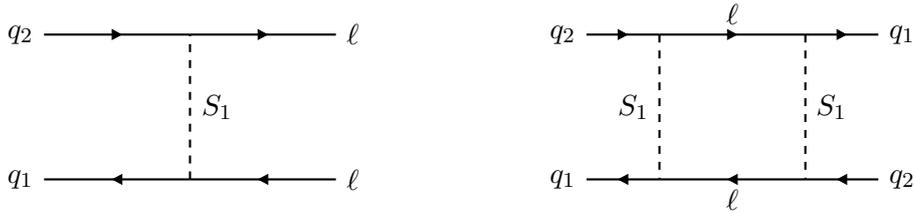
\noindent
A final comment is in order: generating both ${\cal C}_{8gs}^{\rm NP}$ and ${\cal C}_{8gd}^{\rm NP}$, as preferred by $L$-observables taken together with the branching fractions, requires that $S_1$ couples to both first and second generation quarks $q_{1}$ and $q_2$, e.g.
\begin{align}
\mathcal{L}_{S_{1}} &\supset \lambda_{L}^{i} \bar{q}^{ci}_L \epsilon \ell_L^3 S_1\,,
\end{align}
where $i = 1,2$ is a light family flavor index. Because the coupling $\lambda_{L}^{i}$ explicitly breaks the $U(2)_q$ part of the $U(2)^5 \equiv U(2)_q \times U(2)_u \times U(2)_d \times U(2)_\ell \times U(2)_e$ accidental approximate flavor symmetry of the SM, one in general expects strong bounds from FCNC processes if $M_1 \sim $ TeV and $\lambda_{L}^{i} \gtrsim (|V_{td}|,|V_{ts}|)$. Indeed, one can draw a $\Delta F =1 $ diagram giving rise to $K\rightarrow \pi\nu\bar\nu$ at tree-level, as well $\Delta F =2 $ processes at 1-loop level, as shown in \cref{fig:S1FCNC}. Even if $S_1$ does not couple to $d_L$, but only to $s_L$ in the light-family sector, $D-\bar D$ mixing still imposes tight constraints due to the Cabibbo-angle misalignment between the up and down sector.
A simple way to avoid these issues is to promote $S_1 \rightarrow S_1^i$, a doublet of $U(2)_q$. In this case, the coupling with $q_L^i$ need no longer break the $U(2)_q$ symmetry
\begin{align}
\mathcal{L}_{S_{1}} &\supset \lambda_{L} \bar{q}^{ci}_L \epsilon \ell_L^3 S_1^i\,,
\end{align}
forbidding the diagrams in \cref{fig:S1FCNC}  while still allowing both the $s_L$ and $d_L$ chromomagnetic dipoles if the breaking of $U(2)_q$ is shifted to the coupling between $N_R$ and $b_R$. This can be simply understood by the fact that there is now one $S_1$ that couples to each flavor. This is the setup that we will work with for the reminder of the paper.

\subsection{Connection to TeV-scale neutrino mass generation}
The requirement of a TeV-scale RH neutrino forbids a high scale solution for neutrino masses via the usual Type-I seesaw mechanism. Instead, such a scenario naturally points to the \emph{inverse-seesaw mechanism} (ISS) as the origin of small neutrino masses. The mechanism requires the introduction of 3 Dirac singlet fermions $N^\alpha_{L,R}$, with $\alpha=1,2,3$. The relevant Lagrangian is
\begin{align}
\mathcal{L}_{N} &= - \bar \ell_L Y_N \tilde{H} N_R - \bar N_L \hat M_R N_R -\frac{1}{2}  \bar N_L \mu N_L^c \,,
\end{align}
and we can always work in a basis where $\hat M_R$ is a diagonal $3\times 3$ matrix, while $Y_N$ and $\mu$ are arbitrary complex and complex-symmetric $3\times 3$ matrices, respectively. The ISS mechanism explains the smallness of neutrino masses via the smallness of the Majorana mass $\mu$, which is technically natural as it is the only parameter violating lepton number. After EWSB and in the limit of small $\mu$, there are 3 light Majorana states with mass
\begin{equation}
M_{\rm light} \approx \langle H \rangle^2 (Y_N \hat M_R^{-1}) \mu (Y_N \hat M_R^{-1})^T \,,
\end{equation}
corresponding to the active SM neutrinos. Additionally, there are 6 heavy Majorana states that form 3 pseudo-Dirac pairs with masses approximately given by the diagonal of $\hat M_R$. One option is that $\hat M_R = M_R \,\mathbb{1}$, in which case all 3 pseudo-Dirac pairs would lie at the TeV scale. Another possibility is that $\hat M_R$ is very hierarchical. A particularly interesting option, first proposed in Ref.~\cite{Fuentes-Martin:2020pww} and further developed in~\cite{Fuentes-Martin:2022xnb}, 
consists of assuming that $Y_N$ follows a hierarchy similar to that observed in the SM up-quark sector, namely $Y_N \sim V_{\rm CKM}^{\dagger} {\rm diag}(y_u, y_c, y_t)$, while $\hat M_R  \sim {\rm diag}(10^4, 10^2, 1)$ TeV follows an inverse hierarchy. In particular, the low-energy theory consists of only one pseudo-Dirac pair lying at the TeV scale. Integrating out the heavier pseudo-Dirac states, the leading interactions at the TeV scale are
\begin{align}
\mathcal{L}_{N}({\rm TeV}) &\approx - y_{N} \bar \ell_L^{3} \tilde{H} N_R^3 - \hat{M}_R^{(3)} \bar N_L^3 N_R^3 \,,
\end{align}
where we have dropped the Majorana mass $\mu_{33}$ which must be of order $\mu_{33} \sim m_\nu (\hat{M}_R^{(3)}/m_t)^2$ $\sim 1$ eV in order to explain the smallness of the tau neutrino mass. Dropping the Majorana mass is equivalent to treating $N$ as a Dirac state with mass $\hat{M}_R^{(3)}$. We work with this low-energy theory for the rest of the paper and for simplicity we drop the flavor index on $N_R$ and define $\hat{M}_R^{(3)} = M_R$, i.e. we consider one Dirac singlet $N_{L,R}$ with a TeV scale mass.

\subsection{The Model}
We now have all the necessary ingredients to introduce our model to explain the non-leptonic anomalies. It is defined by the following Lagrangian
\begin{equation}
\mathcal{L} = \mathcal{L}_{\rm SM} + \mathcal{L}_{S_{1}}  + \mathcal{L}_{N} + {\rm h.c.} \,,\label{eq:ModelLag}
\end{equation}
with
\begin{align}
\mathcal{L}_{S_{1}} &= \lambda_{L} \bar{q}^{ci}_L \epsilon \ell_L^{3} S_1^{i} + V_{R}^{i} \, \bar{b}_{R}^{c} N_R  S_1^{i} - M_1 S_1^{\dagger i} S_1^i \,, \\
\mathcal{L}_{N} &= - y_{N} \bar \ell_L^{3} \tilde{H} N_R - M_R \bar N_L N_R \,.
\end{align}
The fields $S^i_1=(S_1^{1},S_1^{2})$ are two scalar leptoquarks with quantum numbers $(\bar {\bf 3},{\bf 1})_{1/3}$ under the SM gauge group $SU(3)_c\times SU(2)_L\times U(1)_Y$, arranged as a doublet of the flavor group $U(2)_q$ (where $i=1,2$ is a light family flavor index).
In this model the $U(2)^3 = U(2)_q\times U(2)_u\times U(2)_d$ symmetry in the quark sector is broken only by $V^i_R$, together with the usual breaking from the SM Yukawas that can be written using the $U(2)^3$-breaking spurions $V_q$, $\Delta_u$ and $\Delta_d$,
\begin{equation}
Y_{u,d}=
y_{t,b}
\begin{pmatrix}
\Delta^{ij}_{u,d} && x_{t,b} V^i_q\\
0 && 1
\end{pmatrix},
\label{eq:YukSpurions}
\end{equation}
where we fix the normalization $x_t-x_b=1$.
These spurions have the quantum numbers $V_R\sim ({\bf \bar 2},{\bf 1},{\bf 1})$, $V_q\sim ({\bf 2},{\bf 1},{\bf 1})$, $\Delta_{u}\sim ({\bf 2},{\bf \bar 2},{\bf 1})$ and $\Delta_{d}\sim ({\bf 2},{\bf 1},{\bf \bar 2})$ under $U(2)_q\times U(2)_u\times U(2)_d$, and $x_{t,b}$ measures the amount of alignment between the interaction and mass basis for the third family.
Without loss of generality, we work in the down basis of the 12 sector (i.e. the basis that makes $\Delta_d$ diagonal), but leave $x_t$ and $x_b$ general to avoid unnecesary tunings. The rotation matrices to go from the interaction basis to the mass basis are explicity given in~\cref{app:Rotation}. Furthermore, we take
\begin{equation}
V_{R}^{T} = (-V_{td} / V_{ts}\,, 1) \,\lambda_R^b \,,
\label{eq:VRval}
\end{equation}
in order to obtain ${\cal C}_{8gs} = -{\cal C}_{8gd}$ and we have introduced the coupling $\lambda_R^b$.

The $N_{L,R}$ are total SM singlets carrying lepton number 1. Similarly, the $S_1$ leptoquark can consistently be assigned lepton number -1, so we can write the coupling to $N_R$ but not to $N_L^c$. Integrating out $N$ gives tree-level contributions to the SMEFT Warsaw basis operators $C_{H\ell}^{(1)}$ and $C_{H\ell}^{(3)}$. Including also leading-log running in $y_t$, at the EW scale we get
\begin{align}
[C_{H\ell}^{(1)}]_{33} = -[C_{H\ell}^{(3)}]_{33} =\frac{\theta_\tau^2}{2 v^2}\left[1 + \frac{y_t^2}{16\pi^2} N_c \log\left(\frac{\mu^2_{\rm EW}}{M_R^2} \right)\right]  \,,\label{eq:CHlNR}
\end{align}
where we have defined the mixing angle $\theta_\tau$ between the active tau neutrino and the heavy pseudo-Dirac singlet $N$ as
\begin{equation}
\theta_\tau^2 \equiv \frac{|y_N|^2 v^2}{2 M_R^2} \,,
\label{eq:mixAngle}
\end{equation}
where $v = 246$ GeV is the Higgs vacuum expectation value (VEV). The operators $C_{H\ell}^{(1,3)}$ give important constraints as they modify the couplings of EW gauge bosons to third-family leptons, namely $W\rightarrow \tau \nu_\tau$ and $Z\rightarrow \nu_\tau \nu_\tau$. Combining the EW likelihood provided in~\cite{Breso-Pla:2021qoe} with LFU tests in $\tau$ decays~\cite{Allwicher:2023aql}, we get a constraint on the mixing angle from~\cref{eq:CHlNR} of $\theta_{\tau}\leq 0.05$ at 95\% CL.\footnote{Similar constraints have been found for example in~\cite{Antusch:2014woa,Antusch:2015mia,Coutinho:2019aiy,Crivellin:2020ebi}.} As a point of reference, for $M_R = 2$ TeV this translates into $y_N \leq 0.57$.
In general, we take into account all RGE contributions at leading-log due to $y_t$ and the SM gauge couplings, and we explicitly give the expressions for all WCs relevant for EW precision and LFU tests in $\tau$-decays in Appendix~\ref{app:EWWC}.

\subsubsection{SMEFT dipole matching}
We proceed now to calculate the SMEFT dipole operators relevant for ${\cal C}_{7\gamma}$ and ${\cal C}_{8g}$, as shown in~\cref{fig:FeyDiag}. We define the dipole operators as
\begin{equation}
\mathcal{L}_{\rm SMEFT} \supset C_{dG} \, (g_s \mathcal{O}_{dG}) + C_{dW} \, (g_L \mathcal{O}_{dW}) + C_{dB} \, (g_Y \mathcal{O}_{dB}) \,,
\end{equation}
where $\mathcal{O}_{dG}$, $\mathcal{O}_{dW}$, and $\mathcal{O}_{dB}$ are the operators in the Warsaw basis~\cite{Grzadkowski:2010es}
\begin{equation}
{\cal O}_{dG}= (\bar q_L^i \sigma^{\mu\nu} T^a d_{R}^j ) H G_{\mu\nu}^a \,, \quad {\cal O}_{dW}= (\bar q_L^i \sigma^{\mu\nu} \tau^I d_{R}^j ) H W_{\mu\nu}^I \,,\quad {\cal O}_{dB}= (\bar q_L^i \sigma^{\mu\nu}  d_{R}^j ) H B_{\mu\nu} \,,
\end{equation}
where $T^a$ are the $SU(3)$ generators and $\tau^I$ the Pauli matrices.
Our own computation in \texttt{Package-X}~\cite{Patel:2015tea} cross-checked with the \texttt{Matchete} 1-loop matching software~\cite{Fuentes-Martin:2022jrf} yields the following Wilson coefficients at the matching scale $M_1$
\begin{align}
[C_{dG}]_{i3} &= \frac{1}{16\pi^2} \frac{\lambda_L y_{N} V^{i*}_R}{M_1^2} G(x_R) \,, \\ 
[C_{dW}]_{i3} &= \frac{1}{32\pi^2} \frac{\lambda_L y_{N} V^{i*}_R}{M_{1}^2} W(x_R) \,, \\ 
[C_{dB}]_{i3} &= -\frac{1}{16\pi^2} \frac{\lambda_L y_{N} V^{i*}_R}{M_{1}^2} \bigg[  Y_{S_1}  G(x_R)+ Y_{\ell_L} W(x_R) \bigg]\,, \label{eq418}
\end{align}
where $x_R = M_R / M_1$, and $Y_{S_1} = 1/3$ and $Y_{\ell_L} = -1/2$ are the hypercharges. The loop functions are
\begin{equation}
G(x) = \frac{1-x^4+2x^2 \log x^2}{4(1-x^2)^3} \,, \hspace{15mm} W(x) = \frac{1-x^2+\log x^2}{4(1-x^2)^2} \,,
\end{equation}
which satisfy $G(1) = 1/12$ and $W(1) = -1/8$. In principle, these WCs are calculated in the interaction basis, and therefore in general there is a rotation to move to the down-quark mass basis. This rotation gives an irrelevant contribution to the dipoles that we neglect here. However, in \cref{sec:Constraints} we discuss the impact of this rotation in other observables, where the effect is not negligible.

\subsubsection{SMEFT Running and WET Matching: Extraction of ${\cal C}_{7\gamma}$ and ${\cal C}_{8g}$}
The three dipoles generated at the UV matching scale mix into each other under RGE. Using \texttt{DsixTools}~\cite{Celis:2017hod}, we find the following numerical RGE matrix
\begin{align}
\begin{pmatrix}
[C_{dG}]_{i3} \\
[C_{dB}]_{i3} \\
[C_{dW}]_{i3}
\end{pmatrix}_{\hspace{-1.5mm}\mu_{\rm EW}} 
&= \begin{pmatrix}
0.952 & 0.001 & -0.036 \\
0.016 & 0.932 & -0.016 \\
-0.047 & -0.002 & 0.909
\end{pmatrix}
\begin{pmatrix}
[C_{dG}]_{i3} \\
[C_{dB}]_{i3} \\
[C_{dW}]_{i3}
\end{pmatrix}_{\hspace{-1.5mm}M_1} \,, 
\end{align}
where we have taken the matching scale to be $M_1 = 2$ TeV and $\mu_{\rm EW} = 160$ GeV. The matching condition for photon dipole in terms of the dimensionless WET Wilson coefficients $C_{7\gamma s}$ and $C_{7\gamma d}$ reads
\begin{align}
C_{7\gamma i} &= -\frac{v}{G_F m_b}\frac{4\pi^2}{V_{tb}V_{ti}^*} \left(  [C_{dB}]_{i3} -  [C_{dW}]_{i3} \right)_{\mu_{\rm EW}} \,, 
\end{align}
where $G_F^{-1}= \sqrt{2}v^2$ and we have used $\langle H \rangle = v/\sqrt{2}$. If we neglect the small RGE mixing, we see that only the $Y_{S_1}$ term in~\cref{eq418} (coming from attaching a hypercharge gauge boson to $S_1$ in the loop) contributes to the photon dipole.  Therefore, a good approximate analytic formula is
\begin{align}
C_{7\gamma i} &\approx \frac{v^2}{2m_b}\frac{Y_{S_1}}{V_{tb}V_{ti}^*}\frac{\lambda_L \theta_\tau V_R^{i*}}{M_{1}} \,  x_R \, G(x_R) \,, 
\end{align}
where $\theta_\tau$ is the mixing angle between the active tau neutrino and the heavy pseudo-Dirac singlet $N$ defined in \cref{eq:mixAngle}. The function $x_R\, G(x_R)$ is maximized for $x_R = 1$ (where $N$ and $S_1$ have the same mass), so we work in this limit in what follows. Similarly, for the chromomagnetic dipole, we find 
\begin{equation}
    C_{8gi} \approx -C_{7\gamma i} / Y_{S_1} \,, 
\end{equation}
so before RGE we predict $C_{8gi} = -3C_{7\gamma i}$. 
Including RGE effects, this gets modified to ${C_{8gi} \approx -3.8 C_{7\gamma i}}$ at the EW scale. As we will see, this RG mixing helps to achieve a better accidental cancellation in $b\rightarrow s\gamma$.

\subsection{Computation of the $L$-observables, $B \to X_{s/d}\, \gamma$, and high-$p_T$ constraints}
\begin{figure}
    \centering
    \includegraphics[scale=0.73]{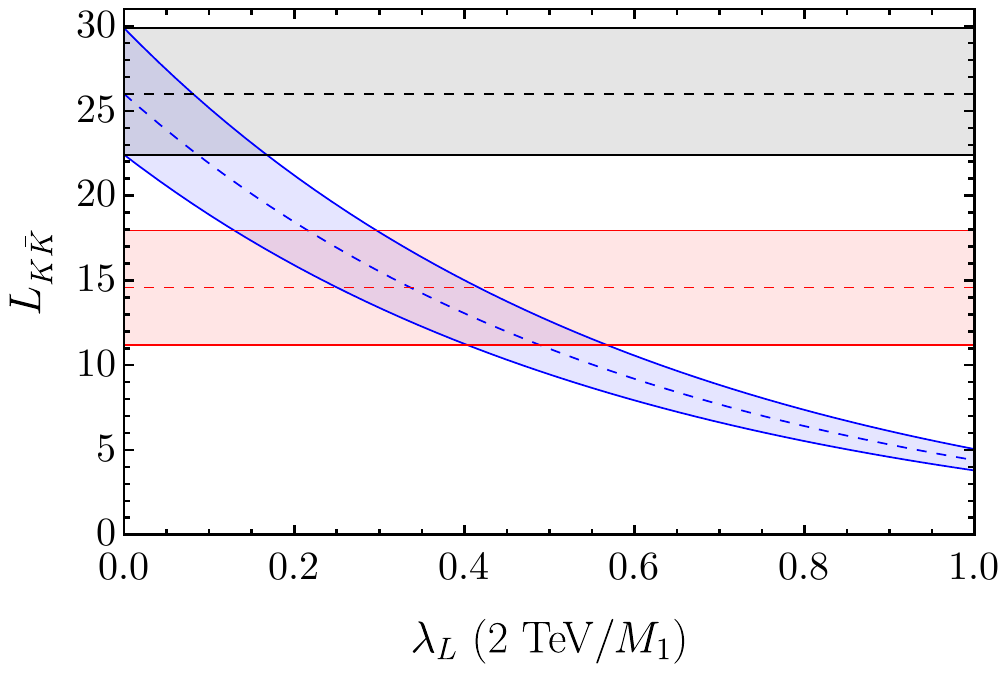}
    \includegraphics[scale=0.73]{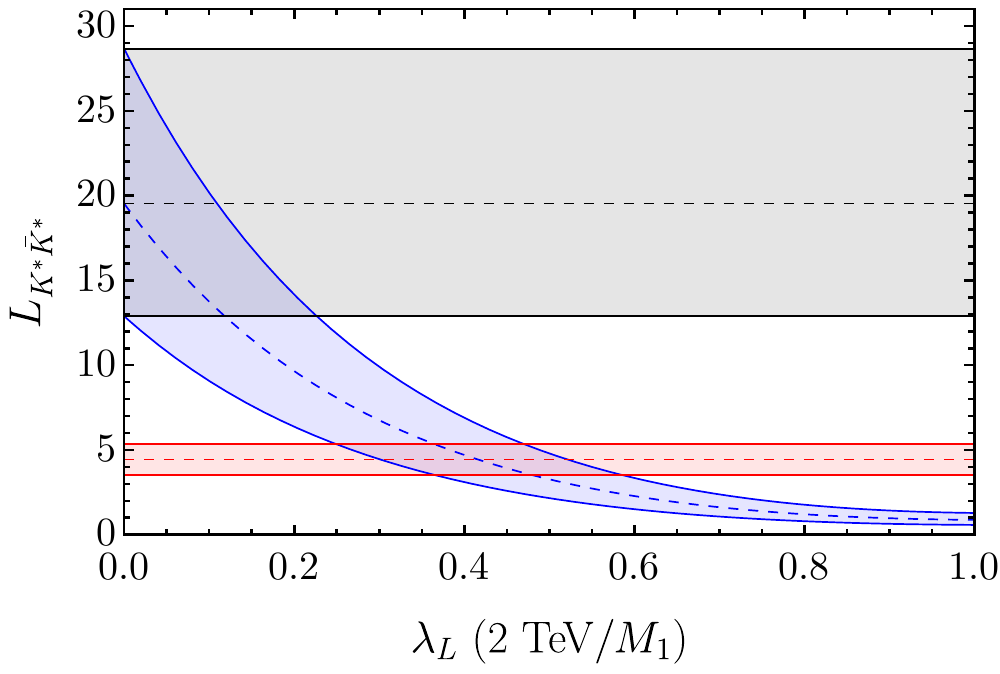}
    \caption{$L$-observables as a function of the coupling $\lambda_L$ normalized to $M_1 = 2$ TeV. The other couplings are fixed as: $\theta_\tau = 0.05$ and $\lambda^b_R = -2$. We have also assumed $C_{8gd} = -C_{8gs}$. Blue: NP theory prediction, Red: Exp. measurement, Black: SM theory prediction.}
    \label{fig:my_label}
\end{figure}
To compute observables such as $L_{K^{(*)}\bar K^{(*)}}$ at $\mu = m_b$, we also need to account for WET running from the EW scale down to $m_b$. Using \texttt{DsixTools}, we find
\begin{equation}
\begin{pmatrix}
C_{7\gamma i} \\
C_{8 gi} 
\end{pmatrix}_{\hspace{-1mm} m_b} 
= \begin{pmatrix}
0.89 & 0.13 \\
0.00 & 0.92
\end{pmatrix}
\begin{pmatrix}
C_{7\gamma i} \\
C_{8 gi} 
\end{pmatrix}_{\hspace{-1mm}\mu_{\rm EW}} \,.
\end{equation}
The expressions for the $L_{K^{(*)}\bar K^{(*)}}$-observables in terms of the NP contributions to the chromomagnetic dipole Wilson coefficients are~\cite{Biswas:2023pyw}
\begin{align}
L_{K\bar K}  = L_{K\bar K}^{\rm SM} &\left[\frac{1+1.23\, C_{8 g s}+ 0.40\, C_{8 g s}^{\,2}}{1+1.34\, C_{8 g d}+ 0.47\, C_{8 g d}^{\,2}}\right]_{m_b} \,, \nonumber \\ 
L_{K^{*} \bar K^{*}}  = L_{K^{*} \bar K^{*}}^{\rm SM}  &\left[ \frac{1+2.62\,C_{8 g s}  + 2.05\, C_{8 g s}^{\,2}}
{1+2.63\,C_{8 g d}  + 2.07\, C_{8 g d}^{\,2}} \right]_{m_b}\,,
\end{align}
where the SM prediction and the experimental values of $L_{K^{(*)} \bar{K}^{(*)}}$ are given in \cref{LSM} and \cref{Lexp}, respectively.
In~\cref{fig:my_label} we present the predictions for the $L$-observables as a function of the relevant coupling parameter $\lambda_L$ normalized to $M_1=2$ TeV. 

Switching now to the theory expressions for $B \to X_{s/d}\, \gamma$ where we follow Ref.~\cite{Misiak:2020vlo}, at $\mu_{\rm EW} = 160$ GeV we have 
\begin{align}
\mathcal{B}_{s\gamma} \times 10^4  & = (3.39 \pm 0.17) - 2.10\, \left(3.93 \, C_{7\gamma s} + C_{8g s}\right)_{\mu_{\rm EW}} 
 \,, \\
\mathcal{B}_{d\gamma} \times 10^4  & = (0.174^{+0.009}_{-0.020}) - 0.09\, \left(3.93 \, C_{7\gamma d} + C_{8g d}\right)_{\mu_{\rm EW}} \,.
\end{align}
These are linearised expressions  that provide an accurate description of the ${\cal B}_{sq}$ working at NNLO in QCD not interpolating
but computing the $m_c$ contribution. One can estimate the uncertainty of the coefficients of the linear term to be of order $5\%$. This estimate is obtained taking into account the combination of two effects \cite{Misiak_private}: i) the effect of the scale variation 2 GeV $\leq \mu \leq$ 5 GeV on these coefficients and ii) half of the overall uncertainty of the SM amplitude, given that these terms are interference terms between SM and BSM.
These expressions supersede those in \cite{Misiak:2015xwa}. One may also consider the impact of quadratic terms.
However, due to the cancellations between the contributions of $C_{7\gamma}$ and  $C_{8g }$ in our model described below, the contribution of the quadratic correction is for a large set of values of $C_{8g }$ estimated to be at the percent level of the central value or below.

Notice that the NP contribution in both cases ($b \to s \gamma$ and $b\to d \gamma$) exactly cancels if $C_{8g}=-3.93 C_{7\gamma}$, which is accidentally very close to our relation at the EW scale $C_{8g}=-3.8 C_{7\gamma}$.
The experimental measurements are~\cite{HFLAV:2019otj,Misiak:2015xwa}
\begin{align} \label{bsg}
\mathcal{B}_{s\gamma}^{\rm Exp} =& (3.32 \pm 0.15)\times 10^{-4}\,,\\
\mathcal{B}_{d\gamma}^{\rm Exp}  =& (0.141 \pm 0.057)\times 10^{-4}\,.
\end{align}
\begin{figure}
    \centering
    \includegraphics[scale=0.59]{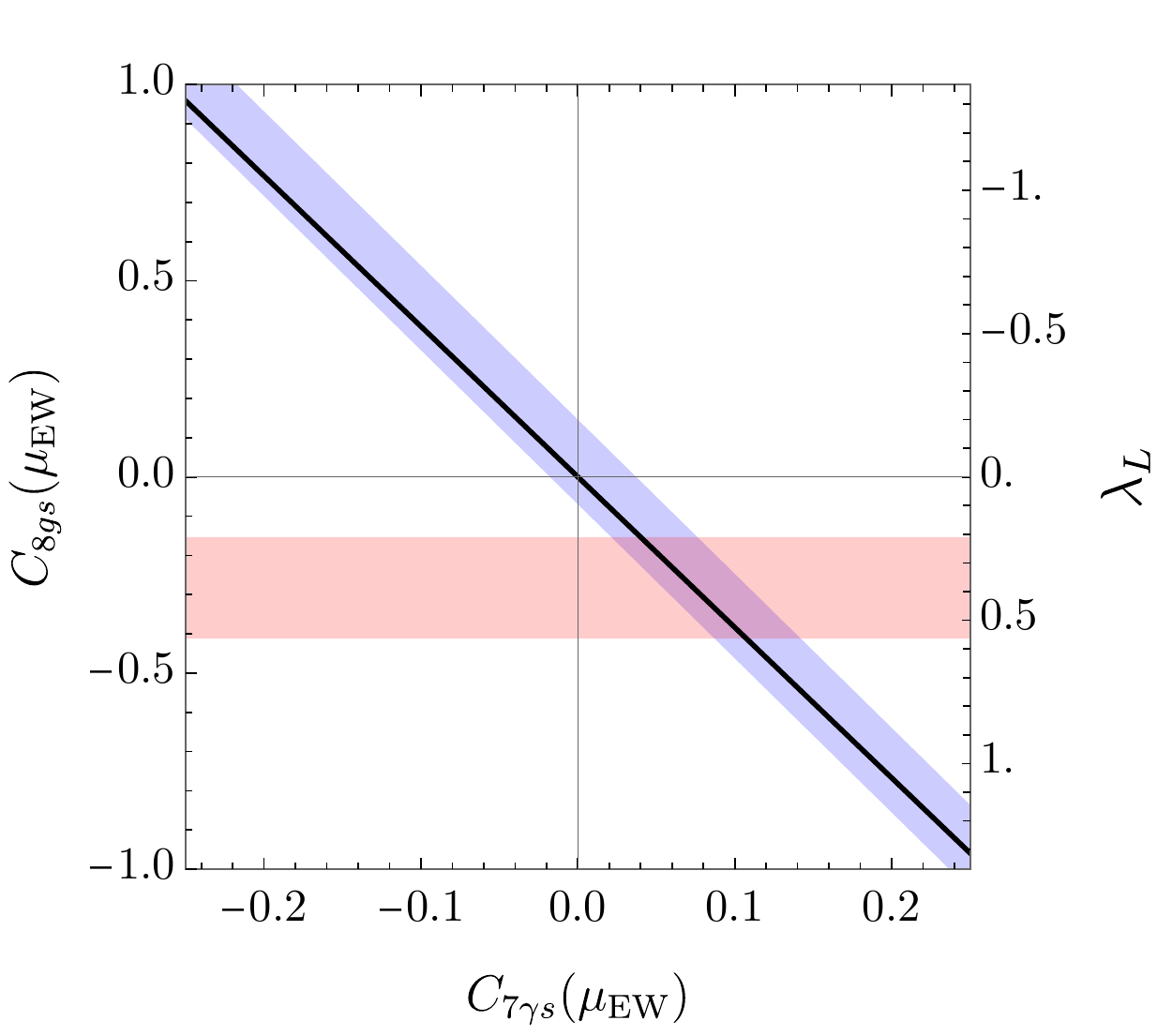}
    \includegraphics[scale=0.59]{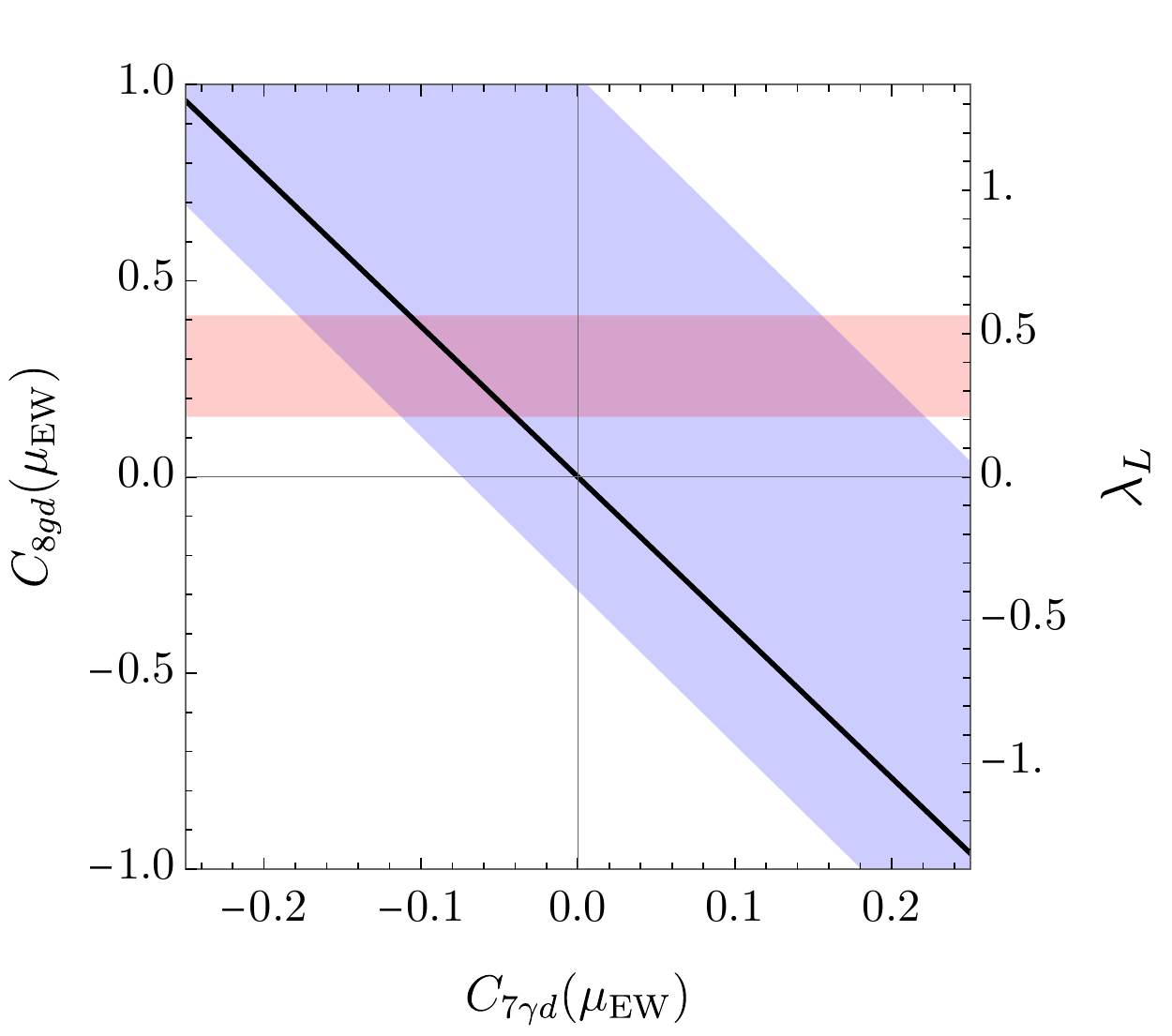}
    \caption{$C_{7\gamma i}$ and $C_{8g i}$ at the EW scale for the model defined in~\cref{eq:ModelLag} (black line). On the left panel we show $C_{7\gamma s}$ and $C_{8g s}$, and on the right panel, $C_{7\gamma d}$ and $C_{8g d}$. In both cases, on the right vertical axis we show the corresponding coupling $\lambda_L$ fixing the other relevant parameters to $M_{S_1}=2\,$TeV, $\lambda^b_R=-2$ and $\theta_{\tau}=0.05$. In blue we show the $\Delta\chi^2=1$ region for
    $\mathcal{B}_{s\gamma}$ (left) and  $\mathcal{B}_{d\gamma}$ (right). In red we show the $\Delta\chi^2=1$ region for
    $L_{KK}$ and $L_{K^{*}K^{*}}$ assuming $C_{7\gamma d}=-C_{7 \gamma s}$ and $C_{8g d}=-C_{8 g s}$. The black line gives the theory prediction in our $S_1$ model of ${C_{8gi} \approx -3.8 C_{7\gamma i}}$ at the EW scale. }  
    \label{fig:C7C8}
\end{figure}
The preferred values for the Wilson coefficients $C_{7\gamma i}$ and $C_{8g i}$ at the EW scale, considering the $L$-observables and the $B \to X_{s/d}\, \gamma$ constraint, are displayed in Fig.\ref{fig:C7C8}.

\subsubsection{Direct searches for $S_1$ leptoquarks at high-$p_T$} 
\label{sec:highPT}

 \begin{figure}[t]
    \centering
    \begin{tabular}{c@{\hskip 0.6cm}c}
         \begin{tikzpicture}[thick,>=stealth,scale=1.2,baseline=-0.5ex]
            \draw[midarrow] (-1.6,0.8) node[left] {$u$} -- (0,0.8)  ;
            \draw[midarrow] (0,0.8) -- (1.6,0.8) node[right] {$\tau^+$};
            \draw[dashed] (0,-0.8) -- (0,0.8); 
            \draw[midarrow] (0,-0.8) -- (-1.6,-0.8) node[left] {$\bar d$};
            \draw[midarrow] (1.6,-0.8) node[right] {$ \nu_\tau$} -- (0,-0.8);
            \node[right] at (0.01,0) {$S_1$};
        \end{tikzpicture}
        \hspace{20mm}
         \begin{tikzpicture}[thick,>=stealth,scale=1.2,baseline=-0.5ex]
            \draw[midarrow] (-1.6,0.8) node[left] {$u$} -- (0,0.8)  ;
            \draw[midarrow] (0,0.8) -- (1.6,0.8) node[right] {$\tau^+$};
            \draw[dashed] (0,-0.8) -- (0,0.8); 
            \draw[midarrow] (0,-0.8) -- (-1.6,-0.8) node[left] {$\bar u$};
            \draw[midarrow] (1.6,-0.8) node[right] {$\tau^-$} -- (0,-0.8);
            \node[right] at (0.01,0) {$S_1$};
        \end{tikzpicture}
    \end{tabular}
    \caption{Diagrams constraining the model at high-$p_T$. Left:  Charged-current mono-$\tau$. Right: Neutral-current di-$\tau$ Drell-Yan.}
    \label{fig:highpTdiag}
\end{figure}
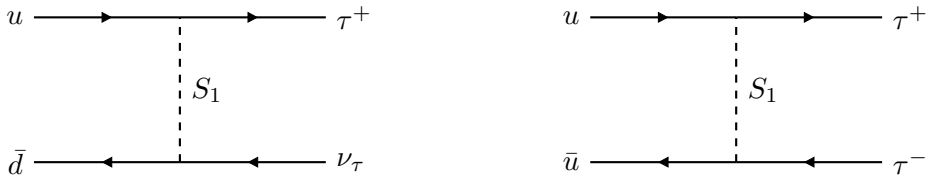

Pure QCD pair production of $S_1$ LQs gives a rather weak lower bound on the LQ mass of $M_1 \gtrsim 1.25$ TeV~\cite{ATLAS:2023vxj}. 
Another interesting recent analysis from CMS~\cite{CMS-PAS-EXO-22-018} studying $S$-channel production of LQs via quark-lepton fusion at the LHC~\cite{Buonocore:2022msy} finds bounds for leptoquarks that in our case imply $\lambda_L <0.9$ at $95\%$ CL for $M_1=2\,$TeV. 

However, a stronger constraint comes from high-mass $\tau\tau$ Drell Yan tails or from mono-$\tau+ \slashed{E}_T$ searches, due to $t$-channel LQ exchange and production via valence quarks, as shown in~\cref{fig:highpTdiag}. We take these constraints into account using the \texttt{HighPT} program~\cite{Allwicher:2022mcg} keeping the full information of the leptoquark propagator.
We generate an estimated event yield for each leptoquark in each bin including the two searches~\cite{ATLAS:2020zms,ATLAS-CONF-2021-025}.
The total estimated event yield per bin is then given by the sum of the two partial event yields for each leptoquark. Notice that there are no interference terms between both leptoquarks when working in the up-basis due to the $U(2)$-symmetry since each leptoquark couples to only one up family.
We can then construct the combined $\tau\tau + \tau\nu$ likelihood as a function of the model parameters.

\subsubsection{Fit to ($\lambda_L,\theta_\tau$) including $L$-observables and all constraints}

\begin{figure}[t]
     \centering
     \includegraphics[scale=0.6]{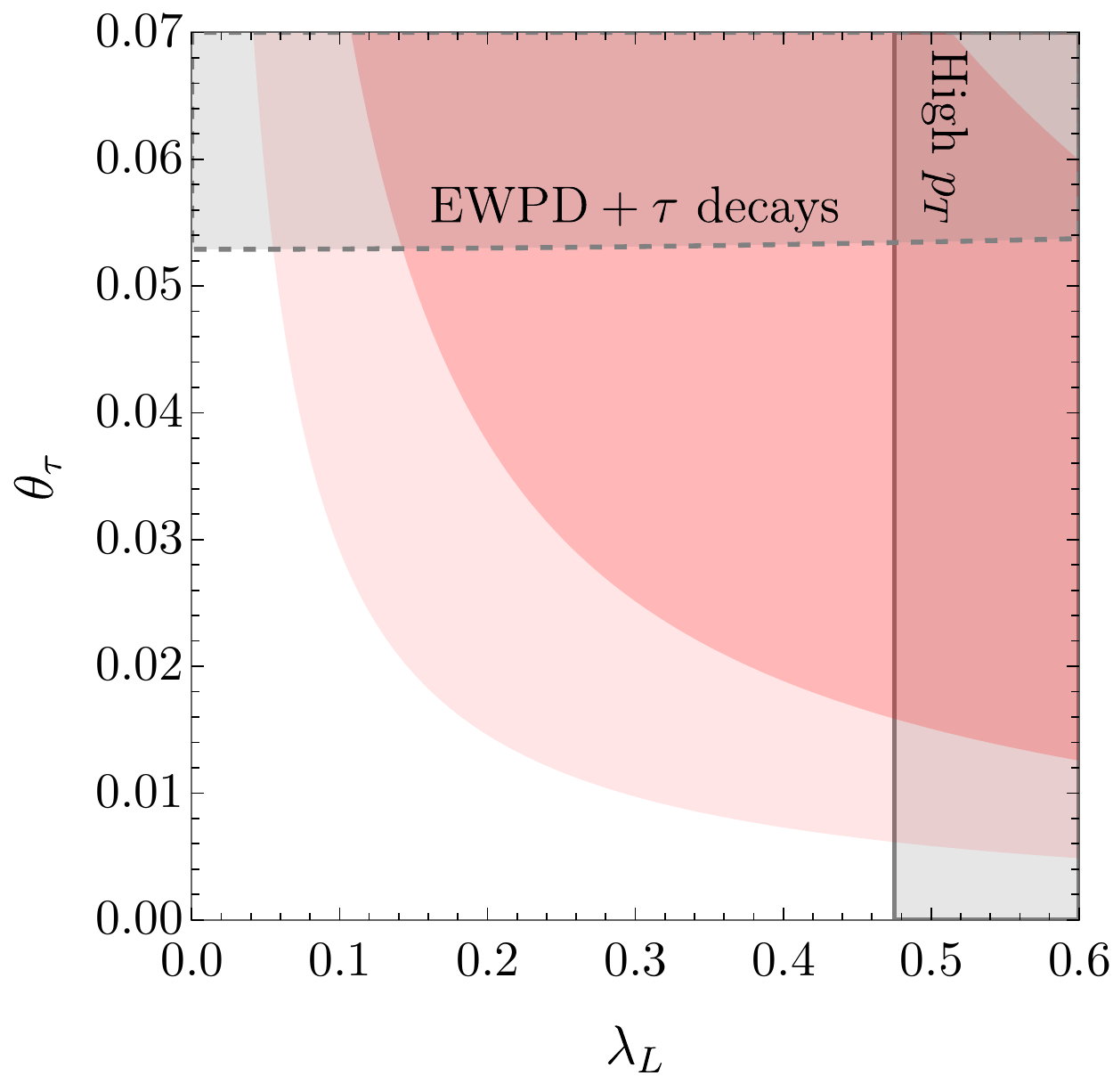} \hspace{5mm}
     \caption{Fit to the $L$-observables
     assuming the relation in \cref{eq:VRval}, giving $C_{7\gamma d}=-C_{7 \gamma s}$ and $C_{8g d}=-C_{8 g s}$. We include   all constraints and fix $M_1=2$\,TeV and $\lambda_R^b=-2$. The red regions are preferred at 1 and 2$\sigma$ by the $L$-observables, while the gray shaded regions are excluded at 95\% CL.}
     \label{fig:GlobalFitThetaLambda}
 \end{figure}

Here we perform a fit taking into account the $L$-observables, EW precision data (EWPD), LFU tests in $\tau$-decays, $\mathcal{B}_{s\gamma}$, $\mathcal{B}_{d\gamma}$, and high-$p_T$ constraints using the likelihood extracted from the \texttt{HighPT} program to determine the allowed ranges for the relevant parameters ($\lambda_L,\theta_\tau$) of our $S_1$ model.
In order to be able to combine the observables, the fit  is done under the approximation of Gaussian distributions, where we symmetrize the error by taking the value in the direction of the SM prediction. 
We fix $M_1=2$\, TeV, a value well above the current exclusion limit from QCD pair-production of LQs. This choice does not make our analysis less general, since changing the LQ mass at this stage is equivalent to re-scaling the couplings. We also fix the coupling $\lambda^b_R=-2$, which determines the normalization of $V_R$. This coupling is not directly constrained by any experimental search, so we choose a sizeable but perturbative value.
The result of the fit is shown in \cref{fig:GlobalFitThetaLambda}.

\section{Connection of the $S_1$ model to charged-current $B$-anomalies ($R_{D^{(*)}}$)}
\label{sec:connect2CC}

To address the $R_{D^{(*)}}$ anomalies in charged-current $B$-decays, we need to switch on new $U(2)$-breaking couplings, $V_L^i$ and $\Delta_R^{ij}$,
\begin{align}
\mathcal{L}_{S_{1}} &=   \lambda_{L} \bar{q}^{ci}_L \epsilon \ell_L^{3} S_1^{i} + V_{R}^{i} \, \bar{b}_{R}^{c} N_R  S_1^{i} + V_L^{i} \,\bar{q}^{c3}_L \epsilon \ell_L^{3} S_1^{i} + \Delta_R^{ij} \, \bar{u}_{R}^{ci} \tau_R  S_1^{j} + {\rm h.c.}\,,\label{eq:S1Lagrangian}
\end{align}
where $V_L^T\equiv (\epsilon_L,1)\lambda_L^b$ is a doublet of $U(2)_q$ and $\Delta_R^{ij}$ is a $({\bf 2},{\bf \bar 2})$ of $U(2)_q\times U(2)_u$.\footnote{Another coupling that could be written at the renormalizable level is the cross-quartic with the Higgs, $\mathcal{L}\supset \lambda^i_H |S_1^i|^2 |H|^2 $. However, the phenomenological impact of this coupling is very suppressed and it is only barely constrained by contributions at the loop level to the $h\gamma\gamma$ and $hgg$ vertices~\cite{Gherardi:2020qhc}.}
Following a minimal $U(2)$-breaking logic, the natural size of $\epsilon_L$ is $ \sim V_{td}/V_{ts}$ while we take 
${\Delta_R^{ij}\equiv \, \lambda_R\frac{m_t}{m_c}\Delta_{u}^{ij}}$  to be aligned with the light Yukawa spurion $\Delta_{u}^{ij}$, defined in \cref{eq:YukSpurions}. This relation could be justified if both spurions have the same origin in a UV completion of the model. 

Integrating out the $S_1$ LQs at tree-level generates the following semi-leptonic operators in the interaction basis
\begin{align}
[C_{lq}^{(1)}]_{3333} = -[C_{lq}^{(3)}]_{3333} &=\frac{V_L^\dagger V_L}{4 M_1^2} \,, \hspace{15mm} [C_{lq}^{(1)}]_{33ij} = -[C_{lq}^{(3)}]_{33ij} =\frac{|\lambda_L|^2}{4 M_1^2} \delta_{ij} \,, \nonumber \\
[C_{lq}^{(1)}]_{33i3} = -[C_{lq}^{(3)}]_{33i3} &=\frac{\lambda_L^* V_L^i}{4 M_1^2}\,, \hspace{15mm} 
[C_{lequ}^{(1)}]_{333i} = - 4[C_{lequ}^{(3)}]_{333i} = \frac{[V_L^\dagger \Delta_{R}]_i }{2 M_1^2} \,,
\nonumber \\
[C_{lequ}^{(1)}]_{33ij} = - 4[C_{lequ}^{(3)}]_{33ij} &= \frac{\lambda_L^* \Delta_{R}^{ij}  }{2 M_1^2} \,,\hspace{14mm}  [C_{eu}]_{33ij} = \frac{[\Delta_{R}^\dagger \Delta_{R}]_{ij}  }{2 M_1^2} \,.
\label{eq:treeSemiLep}
\end{align}
Relevant for $b\rightarrow c\tau \nu$ transitions are $[C_{lq}^{(3)}]_{33i3}$ and $[C_{lequ}^{(1,3)}]_{333i}$. In particular, the expressions for $R_{D,D^*,\Lambda_c}$ normalized to their respective SM values 
read~\cite{Iguro:2018vqb,Becirevic:2022bev}
\begin{align}
\frac{R_D}{R_D^{\rm SM}} &=  |1+\CV|^2 +1.49\,  {\rm Re}\big[(1+\CV) \CSs \big] +1.14\, {\rm Re}\big[(1+\CV) \CTs\big] \nonumber \\
&+  1.02 |\CS|^2 + 0.9 |\CT|^2 \,, \\
\frac{R_{D^*}}{R_{D^*}^{\rm SM}} &=  |1+\CV|^2 -0.11\,  {\rm Re}\big[(1+\CV) \CSs \big] -5.12\, {\rm Re}\big[(1+\CV) \CTs \big] \nonumber\\
&+ 0.04 |\CS|^2 + 16.07 |\CT|^2 \,, \\
\frac{R_{\Lambda_c}}{R_{\Lambda_c}^{\rm SM}} &=  |1+\CV|^2 +0.34 \,  {\rm Re}\big[\CS+\CV\CSs  \big] -3.10\, {\rm Re}\big[(1+\CV^*) \CT \big] \nonumber\\
&+ 0.34 |\CS|^2 + 10.43 |\CT|^2 \,,
\end{align}
where the Wilson coefficients are understood to be evaluated at the scale $m_b$. In terms of the SMEFT Wilson coefficients at the matching scale in the interaction basis, we find
\begin{align}
\CV (m_b) &= -\eta_V \frac{v^2}{V_{cb}} \left(\sum_{i=1}^{2} V_{2i}[C_{lq}^{(3)}]_{33i3}
+x_t V_{cb} [C_{lq}^{(3)}]_{3333}
\right),\label{eq:CVc}\\
\CS (m_b)&= -\eta_S \frac{v^2}{2V_{cb}}[C_{lequ}^{(1)}]_{3332}^*\, ,\label{eq:CSc}\\
\CT (m_b) &= -\eta_T \frac{v^2}{2V_{cb}}[C_{lequ}^{(3)}]_{3332}^* \,,\label{eq:CTc}
\end{align}
where the $\eta$ parameters take care of the running from $M_1 = 2$ TeV to the scale $m_b$. Using \texttt{DsixTools}, we find $\eta_V\approx 1$, $\eta_S \approx 1.7$, and $\eta_T \approx 0.9$. 
The contributions $[C_{lq}^{(3)}]_{33\alpha 3}$ to $R_{D^{(*)}}$ scale with the model parameters as $[C_{lq}^{(3)}]_{3333} \propto x_t V_{cb} V_L^{\dagger} V_L$, $[C_{lq}^{(3)}]_{3323} \propto V_{cs} \lambda_L^* \lambda_L^b$ and $[C_{lq}^{(3)}]_{3313} \propto V_{cd} \lambda_L^*  \epsilon_L$, so we see that $R_{D^{(*)}}$ is connected to the non-leptonic dipoles via the coupling $\lambda_L$, which contributes dominantly via $[C_{lq}^{(3)}]_{3323}$.

\subsection{Constraints connected to $R_{D^{(*)}}$ and the non-leptonic puzzle }
\label{sec:Constraints}

\begin{figure}[h!]
    \centering
    \begin{tabular}{c@{\hskip 0.6cm}c}
         \begin{tikzpicture}[thick,>=stealth,scale=1.2,baseline=-0.5ex]
            \draw[midarrow] (-1.6,0.8) node[left] {$q^c_i$} -- (0,0.8)  ;
            \draw[midarrow] (0,0.8) -- (1.6,0.8) node[right] {$\ell_3$};
            \draw[dashed] (0,-0.8) -- (0,0.8); 
            \draw[midarrow] (0,-0.8) -- (-1.6,-0.8) node[left] {$q^c_3$};
            \draw[midarrow] (1.6,-0.8) node[right] {$\ell_3$} -- (0,-0.8);
            \node[right] at (0.01,0) {$S_1^i$};
            \node[right] at (-0.25,-1.1)  {$V_L^i$};
            \node[right] at (-0.3,1.1) {$\lambda_L^*$};
        \end{tikzpicture}
        \hspace{20mm}
        \begin{tikzpicture}[thick,>=stealth,scale=1.2,baseline=-0.5ex]
            \draw[midarrow] (-1.6,0.8) node[left] {$q^c_{i}$} -- (-0.8,0.8);
            \draw[midarrow] (-0.8,0.8) -- (0.8,0.8);
            \draw[dashed] (0.8,0.8) -- (0.8,-0.8);
            \draw[midarrow] (0.8,-0.8) -- (-0.8,-0.8);
            \draw[midarrow] (-0.8,-0.8) -- (-1.6,-0.8) node[left] {$q^c_{3}$};
            \draw[dashed] (-0.8,0.8) -- (-0.8,-0.8);
            \draw[midarrow] (0.8,0.8) -- (1.6,0.8) node[right] {$q^c_{3}$};
            \draw[midarrow] (1.6,-0.8) node[right] {$q^c_{i}$} -- (0.8,-0.8);
            \node[above] at (0,0.8) {$\ell_3$};
            \node[below] at (0,-0.8) {$\ell_3$};
            \node[right] at (0.8,0) {$S_1^i$};
            \node[left] at (-0.8,0) {$S_1^i$};
            \node[right] at (-1.1,1.1) {$\lambda_L^*$};
            \node[right] at (0.5,-1.1) {$\lambda_L^*$};
            \node[right] at (0.5,1.1) {$V_L^i$};
            \node[right] at (-1.1,-1.1) {$V_L^i$};
        \end{tikzpicture}
    \end{tabular}
    \caption{FCNC diagrams induced when the $S_1$ model solves both the non-leptonic puzzle and the charged-current $B$-anomalies (for non-leptonic puzzle only: $V_L \rightarrow 0$). Left: Tree-level contribution to $B\rightarrow K (\pi) \nu\bar\nu$. Right: One-loop contribution to $B_{s,d}$-$\bar B_{s,d}$ mixing.}
    \label{fig:S1FCNCwRD}
\end{figure}
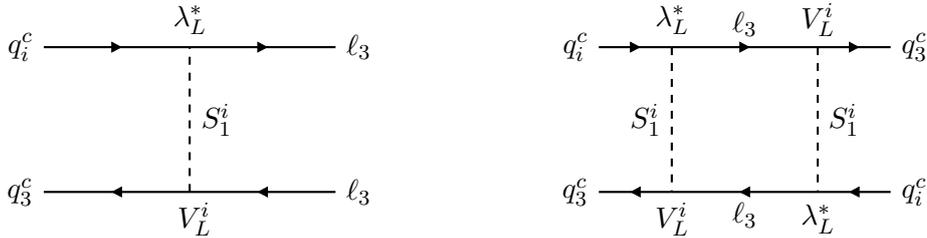
 As we have just seen, the dipole diagram responsible for solving the non-leptonic puzzle is connected to effects in $R_{D^{(*)}}$ via $[C_{lq}^{(3)}]_{3323}$. However, switching on the new couplings $V_L$ and $\Delta_R$ that break the $U(2)$ flavor symmetry also introduces several new constraints. Still, we will see that the $U(2)$ flavor symmetry of the model protects all FCNCs if the bounds on $b\to s$ transitions are passed. This is a common feature of models with Minimal Flavor Violation (MFV)~\cite{DAmbrosio:2002vsn,EuropeanStrategyforParticlePhysicsPreparatoryGroup:2019qin} or minimally-broken $U(2)$~\cite{Davighi:2023evx}.

\subsubsection{$\Delta F=1$ processes} 

The relation $[C_{lq}^{(3)}]_{33i3} = -[C_{lq}^{(1)}]_{33i3}$ at tree-level, depicted in~\cref{fig:S1FCNCwRD} (left), leads unavoidably to large contributions to the $\Delta F = 1$ processes $B\rightarrow K^{(*)}\nu\bar\nu$. They are given by~\cite{Gherardi:2020qhc,Fuentes-Martin:2020hvc}
\begin{align}
R_{K^{(*)}}^{\nu} =&\frac{{\rm Br}(B\to K^{(*)}\nu\nu)}{{\rm Br}(B\to K^{(*)}\nu\nu)_{\rm SM}}=
\frac{2}{3}+\frac{1}{3}\left|1+\frac{C_{sb,\tau}}{C^{\rm SM}_{sb,\tau}}\right|^2,
\end{align}
where $C_{sb,\tau}^{\rm SM} = -1.48 \pm 0.01$ \cite{Buchalla:1998ba}. Defining $\Delta C^\tau_{ij} = [C_{lq}^{(1)}-C_{lq}^{(3)}]_{33ij}$ we can write
\begin{align}
C_{sb,\tau} &=\frac{v^2} {2} \frac{2\pi}{\alpha_w} \frac{1}{V_{tb}V^{*}_{ts}}\Big[\Delta C^\tau_{23}  -x_b V^{*}_{ts}\left(\Delta C^\tau_{33} - \Delta C^\tau_{22} \right) \Big] \approx 
\frac{v^2} {2M_1^2} \frac{\pi}{\alpha_w} \frac{\lambda_L^* \lambda_L^b}{V_{tb}V^{*}_{ts}}
, 
\end{align}
with $\alpha_w=g_L^2/4\pi$. In the approximate equality we have neglected the $x_b$ contribution which is always $V^{*}_{ts}$ suppressed.
The current experimental limits are $R_K^{\nu}=2.4\pm 0.9$ and $R_{K^*}^{\nu}<3.2$ ($95\%$ CL). Combining both measurements and asuming real couplings we obtain 
\begin{equation}
-\left( \frac{1}{3.5\,{\rm TeV}} \right)^2\lesssim\frac{\lambda_L \lambda^b_L}{M_1^2} \lesssim \left( \frac{1}{5\,{\rm TeV}} \right)^2~~~(95\%\,{\rm C.L.}).\label{eq:limitBKnunu}
\end{equation}
Passing this bound imposes a limit on the product $|\lambda_L \lambda_L^b|$, meaning we also
need a sizeable contribution coming from $C_{S_L}$ and $C_T$ if we want to obtain the central values of $R_{D^{(*)}}$.

Likewise, contributions to $B\rightarrow \pi\nu\bar\nu$ are also expected. Both the NP contribution and the SM prediction will have a similar suppression with respect to $B\rightarrow K^{(*)}\nu\bar\nu$ ($\epsilon_L$ and $V_{td}/V_{ts}$ respectively), so the relative impact of NP in both observables is similar. However, the experimental limit on $R_{\pi}^{\nu}={\rm Br}(B\to \pi\nu\nu)/{\rm Br}(B\to \pi\nu\nu)_{\rm SM}$ is one order of magnitude weaker than $R_{K^{(*)}}^{\nu}$ and therefore it is automatically satisfied provided~\cref{eq:limitBKnunu} is satisfied.

In the light-family sector, although $K^+\to \pi^+ \nu\bar \nu$ is protected to some extent by the $U(2)$ symmetry, the spurions $V_L$ and $x_b V_q$ can give contributions.
The LEFT operator
\begin{equation}
\mathcal{L}_{\rm LEFT} \supset C_{ds,\tau} (\bar d_L \gamma_{\mu} s_L)(\bar \nu_{\tau}^{}\gamma^{\mu}\nu_{\tau}) \,, 
\end{equation}
receives NP contributions that at leading order are
\begin{equation}
C_{ds,\tau}=- \frac{x_b}{2M_1^2}(V_{td}^* \lambda_L\lambda_L^{b\,*} + V_{ts} \lambda_L^* \lambda_L^b\epsilon_L)
-\frac{|\lambda_L|^2|\lambda_L^b|^2}{64 \pi^2 M_1^2}\epsilon_L.
\end{equation}
The first term comes from the tree-level generated Wilson coefficients $[C^{(i)}_{lq}]_{33i3}\propto V^i_L$ after being rotated to the down basis, and the second term is generated via a one-loop box diagram.
Following~\cite{Crosas:2022quq}, we find that the limits on $K^+\to \pi^+ \nu\bar \nu$ impose
\begin{equation}
-\left(\frac{1}{50\,{\rm TeV}}\right)^2 \lesssim C_{ds,\tau} \lesssim \left(\frac{1}{80\,{\rm TeV}}\right)^2~~~~~(95\%~\rm{CL}).
\end{equation}
The loop-generated contribution is safely inside this range for couplings $\sim 0.5$ and TeV scale masses (in fact, $B_s$-mixing imposes a stronger constraint on the same combination of couplings and masses, as we will see below). On the other hand, the tree-level contribution roughly translates into the bounds of~\cref{eq:limitBKnunu} if $x_b=-1$, which corresponds to third-family up-alignment. Therefore, although this contribution is under control if $B\to K^{(*)}\nu\bar \nu$ is, some mild down-alignment $|x_b|<|x_t|$ could be helpful to suppress it. However, it could fall within the sensitivity of the NA62 experiment~\cite{NA62:2021zjw}, which aims to measure $\mathcal{B}(K^+\to \pi^+ \nu\bar \nu)$ at the $O(10\%)$ level~\cite{NA62:2022hqi} (and perhaps 5\% ultimately~\cite{NA62KLEVER:2022nea}).

\subsubsection{$\Delta F=2$ processes} 

Integrating out the leptoquarks generates at one loop $4$-quark operators in SMEFT~\cite{Gherardi:2020qhc}  (see~\cref{fig:S1FCNCwRD} right),
\begin{equation}
\mathcal{L}_{\rm SMEFT} \supset -\frac{1}{256 \pi^2 M_1^2}
\left (\sum_{\alpha,\beta=1}^3\Lambda^{\alpha \beta} \bar q_L^{\alpha} \gamma_{\mu} q_L^{\beta}
\right)^2-
\frac{1}{256 \pi^2 M_1^2}
\left (\sum_{\alpha,\beta=1}^3\Lambda^{\alpha \beta} \bar q_L^{\alpha} \tau_a \gamma_{\mu} q_L^{\beta}
\right)^2
\end{equation}
where $\tau_a$ are the $SU(2)_L$ Pauli matrices, and in the interaction basis,
\begin{equation}
\Lambda=\begin{pmatrix}
|\lambda_L|^2 \delta^{ij} && \lambda_L^* V_L \\
\lambda_L V_{L}^{\dagger} && V_{L}^{\dagger} V_L\\
\end{pmatrix}.
\end{equation}
After rotating to the mass basis, we obtain the $\Delta F=2$ operators
\begin{equation}
\mathcal{L}_{\rm LEFT} \supset -C^1_{B_s} (\bar s_L \gamma_{\mu} b_L)^2-
C^1_{B_d} (\bar d_L \gamma_{\mu} b_L)^2-C^1_{D} (\bar u_L \gamma_{\mu} c_L)^2-C^1_{K} (\bar d_L \gamma_{\mu} s_L)^2.
\end{equation}
The leading contribution to these Wilson coefficients is
\begin{align}
C_{B_s}^1=&\frac{\lambda_L^{b\,2}\lambda_L^{*\,2}}{128\pi^2 M_1^2},\phantom{\epsilon_L^2}~~~~~
C_{K}^1=\frac{x_b^2 (V_{td}^* \lambda_L^{b\,*}\lambda_L +V_{ts}\epsilon_L \lambda_L^{b}\lambda_L^* )^2}{128\pi^2 M_1^2},\\
C_{B_d}^1=& \epsilon_L^2 C_{B_s}^1,~~~~~
C_{D}^1=\frac{x_t^2 
(V_{ub}\lambda_L^{b\,*}\lambda_L + V_{cb}^*(V_{us}+\epsilon_L)\lambda_L^{b}\lambda_L^* )^2}
{128\pi^2 M_1^2},
\end{align}
that should be compared to the $95\%$ CL limits given in~\cite{UTfit:2007eik,FlavConstraints}.
Passing the bound from $B_s$-mixing imposes
\begin{equation}
\frac{|\lambda_L^b \lambda_L|^2}{M_1^2} \lesssim \left(\frac{1}{6.5\,{\rm TeV}}\right)^2~~~~~(95\%~\rm{CL})\,,\label{eq:BsMixing}
\end{equation}
and $B_d$-mixing gives a similar bound once we take into account the suppression from $\epsilon_L$ which is order $V_{td}/V_{ts}$.

Although the bounds on the $\Delta F=2$ processes in the light sector are stronger, the $U(2)$ symmetry of the model protects them, so they only receive contributions suppressed by $V_{td}^2$. 
Their exact contribution depends on the precise values of $\epsilon_L$ and $x_{t,b}$, but in the worse case scenario, the strongest constraint (coming from ${\rm Im} C^1_{K}$) gives a similar bound to~\cref{eq:BsMixing}. Analogously to $K^+\to \pi^+\nu\bar \nu$, some amount of third-family down alignment, $|x_b|<|x_t|$, could be preferred to suppress this contribution.

\subsubsection{Other charged-current transitions} 

Besides $R_{D^{(*)}}$, contributions to other charged-current transitions are expected. They include $B_{c,u}\to \tau\nu$ decays, mainly mediated through $[C_{lq}^{(3)}]_{33i3}$ and $[C_{lequ}^{(1,3)}]_{33i3}$, and $D^+_{(s)}\to \tau \nu$ decays, mediated through $[C_{lq}^{(3)}]_{33ij}$ and $[C_{lequ}^{(1,3)}]_{33ij}$. Precise expressions for these observables are given in~\cref{app:ChargeCurrent}.
Among these, the most relevant ones are $B_u\to \tau \nu$ and $D_s\to \tau \nu$, but anyway they are weaker than the other constraints.
In particular, $B_u\to \tau \nu$ imposes
\begin{equation}
\left|\frac{\lambda_L \lambda_L^b}{M_1^2}\right| \lesssim 
\left(\frac{1}{2\,{\rm TeV}} \right)^2 ~~~(95\%~{\rm C.L.}),
\end{equation}
which is automatically satisfied if the constraint from $B\to K^{(*)}\nu\nu$ in~\cref{eq:limitBKnunu} is passed.
At the loop level, the new states can also potentially contribute to charge-current transitions involving the light-family leptons, affecting the experimental determination of $V_{ud}$ or $V_{us}$~\cite{Coutinho:2019aiy,Crivellin:2020ebi}. We have checked that these contributions to the CKM elements cancel at the leading log order, and if any, higher order contributions are estimated to be at least two orders of magnitude below the experimental error of these CKM elements.

\subsubsection{Neutron electric dipole moment}

Although we take all the NP couplings to be real to maximize the impact of NP into the anomalies, possible imaginary parts could give unwanted contributions to the neutron electric dipole moment (EDM).
Electric and chromo-electric dipole moments of the up and down quarks,
\begin{equation}
\mathcal{L} \supset - \sum_{q=u,d}\left(\frac{i}{2} d_q \bar q_L \sigma^{\mu\nu} q_R F_{\mu\nu}
+\frac{i}{2} \tilde d_q \bar q_L \sigma^{\mu\nu} T_a q_R  \,g_s G^a_{\mu\nu} \right)+ {\rm h.c.}
\end{equation}
contribute to the neutron EDM $d_n$. Using QCD sum rules, one can estimate
that this contribution is~\cite{Pospelov:2000bw}
\begin{equation}
d_n = (1\pm 0.5)\left[ 1.4( d_d -0.25 d_u ) + 1.1 e (\tilde d_d +0.5 \tilde d_u)\right].\label{eq:dnQCDSumRules}
\end{equation}
Similar contributions have been found using lattice QCD~\cite{Gupta:2018lvp}.
For the down-quark dipole, we get that after rotating to the mass basis, the leading contributions come from one-loop matching pieces,
\begin{align}
 d_d \sim   \frac{e\,x_b |V_{td}|^2}{288 \,\pi^2}\frac{y_d y_N}{y_b}{\rm Im}(\lambda_L \lambda^{b*}_R/V_{ts}^*) \frac{v}{\sqrt{2}\,M_1^2},~~~~
\tilde d_d \sim  -\frac{3}{e} d_d,
\end{align}
where we are neglecting the running. Even for $O(1)$ imaginary parts of the couplings, assuming TeV masses for $S_1$ and $N$, they are more than one order of magnitude below the experimental bound $|d_n| < 5.5 \cdot 10^{-13}\,{\rm GeV}^{-1}$ ($95\%\,$C.L.)~\cite{EuropeanStrategyforParticlePhysicsPreparatoryGroup:2019qin,Chupp:2017rkp}. 
The up-quark contribution is dominated by the running of $[C^{(3)}_{lequ}]_{3311}\propto \Delta^{11}_R =\lambda_R \, y_{u}/y_c $ into the up-quark EDM,
\begin{equation}
d_u\sim  -\frac{e}{8\pi^2}  {\rm Im}(\lambda_L \lambda_R^*) \frac{y_{\tau} y_u}{y_c}\frac{v }{\sqrt{2}\,M_1^2} \log\left( \frac{m_n}{M_1} \right),
\end{equation}
where $m_n$ is the neutron mass. Taking $M_1= 2\,{\rm TeV}$ and using the central value of~\cref{eq:dnQCDSumRules} the neutron EDM bound implies
\begin{equation}
|{\rm Im}(\lambda_L \lambda_R^*)| < 0.06,
\end{equation}
a constraint not difficult to satisfy if the NP is close to the CP conserving limit.

\subsubsection{EWPT and LFU tests in $\tau$-decays}

When $V_L\neq 0$, the tree-level semi-leptonic operators $[C_{lq}^{(1,3)}]_{3333}$ induced by $S_1$ involving all third-family quarks give additional contributions to $[C_{H\ell}^{(1,3)}]_{33}$ at one-loop due to $y_t$ running:
\begin{align}
\Delta[C_{H\ell}^{(1)}]_{33} = &  \frac{N_c y_t^2}{16\pi^2} [C_{\ell q}^{(1)}]_{3333}\log\left(\frac{\mu_{\rm EW}^2}{M_1^2} \right) \,, \\
\Delta[C_{H\ell}^{(3)}]_{33} =   -&\frac{N_c y_t^2}{16\pi^2} [C_{\ell q}^{(3)}]_{3333}\log\left(\frac{\mu_{\rm EW}^2}{M_1^2} \right) \,.
\end{align} These new contributions correct the $Z\rightarrow \tau \tau$ vertex, in addition to $W\rightarrow \tau \nu_\tau$.
In addition, we include all contributions at leading-log running due to $y_t$ and the SM gauge couplings. We give the expressions for all WC's relevant for the EWPO in Appendix~\ref{app:EWWC}. With them, we construct a likelihood combining the EW likelihood provided in~\cite{Breso-Pla:2021qoe} with LFU tests in $\tau$-decays~\cite{Allwicher:2023aql}.

\subsection{Combined Global Fit: $b\rightarrow c\tau\nu$ and $L_{K^{(*)}\bar K^{(*)}}$-observables }

We next perform a global fit taking into account $b\rightarrow c\tau\nu$ and $L_{K^{(*)}\bar K^{(*)}}$-observables as well as all relevant constraints. In particular, we consider:
\begin{itemize}
    \item The optimized non-leptonic observables $L_{K^*\bar{K}^*}$ and $L_{K\bar{K}}$.
    \item LFU tests in $b\to c\tau \nu$ transitions, including the ratios $R_D$, $R_{D^*}$ and $R_{\Lambda_c}$.
    \item Flavor bounds: $B\to K^{(*)}\nu \bar{\nu}$, $\Delta m_{B_s}$, $b\to s(d)\gamma$, $D_s\to \tau \nu$.
    \item EW precision data and LFU tests in $\tau$-decays.
    \item Constraints from $pp\rightarrow \ell \ell (\ell\nu)$ at high-$p_T$ (see \cref{sec:highPT}).
\end{itemize}
For $b\rightarrow c\tau\nu$ transitions, we use the experimental averages and SM theory predictions given in Ref.~\cite{Aebischer:2022oqe}. The parameters $x_{t,b}$ and $\epsilon_L$ have very little impact on the fit because, as we have shown in the previous section, flavor observables involving the first family, such as $B\to \pi \nu \bar{\nu}$, $K\to \pi \nu \bar{\nu}$, $\Delta m_{B_d}$, meson mixing in the light sector, and $B_u \to \tau \nu$ remain inside their bounds provided we have control on $b\to s$ transitions, especially if we have a mild third-family down alignment, $|x_b|<|x_t|$. For this reason, we do not include them in the fit. Likewise, the neutron EDM does not give any constraint because we assume NP couplings to be real. Additionally, we fix $M_1 = 2$ TeV, $\lambda_R^b = -2$, and assume $C_{7\gamma d}=-C_{7 \gamma s}$ and $C_{8g d}=-C_{8 g s}$, as in all previous plots. We then construct a global likelihood involving the following four parameters controlling all the relevant phenomenology:
\begin{equation}
\theta_{\tau}, \,\, \lambda_L, \,\, \lambda_L^b, \,\, \lambda_R \,.
\end{equation}
\begin{table}[t]
    \centering
    \renewcommand{\arraystretch}{1.4}
    \begin{tabular}{c|c|c|c|c}
         Observable & $L_{KK}$ & $L_{K^*K^*}$ & $R_{D}$ & $R_{D^*}$  \\\hline\hline
         $P_{\rm SM}$ & 2.32 & 2.25 & -1.98 & -2.15 \\
         $P_{\rm BFP}$ & 0.90 & 0.87 & -0.34 & -0.73 \\
    \end{tabular}
    \caption{Pull of $L_{K^{(*)}K^{(*)}}$ and $R_{D^{(*)}}$ in the SM and the BFP of our model.}
    \label{tab:obsPulls}
\end{table}
We find the best fit point (BFP) to be $\theta_{\tau}= 0.034\pm 0.009$, $\lambda_L=0.27\pm 0.05$, $ \lambda_L^b=0.38\pm 0.07$ and $\lambda_R=-1.6\pm 0.3$, corresponding to $\Delta \chi^2=\chi^2_{\rm SM}- \chi^2_{\rm BFP}=19.5$.
In \cref{tab:obsPulls}, we show the pull of $R_{D^{(*)}}$ and $L_{K^{(*)}\bar{K}^{(*)}}$ observables in the SM and the BFP, where we have defined the pull of an observable $O$ in a theoretical model $M$ as
\begin{equation}
P_M=\frac{O_M-O_{\rm Exp}}{\sqrt{\sigma_{\rm Exp}^2+\sigma_{\rm Th}^2}},
\end{equation}
where $\sigma_{\rm Exp(Th)}$ is the experimental (theoretical) uncertainty for the given observable $O$. We observe a significant improvement with respect to the tensions present in the SM.\footnote{The tensions we find in $L_{K^{(*)}\bar{K}^{(*)}}$ for the SM are slightly smaller than the ones found in~\cite{Alguero:2020xca,Biswas:2023pyw} due to the Gaussian approximation we are using.}
\begin{figure}[t]
    \centering
    \includegraphics[height=0.49\textwidth]{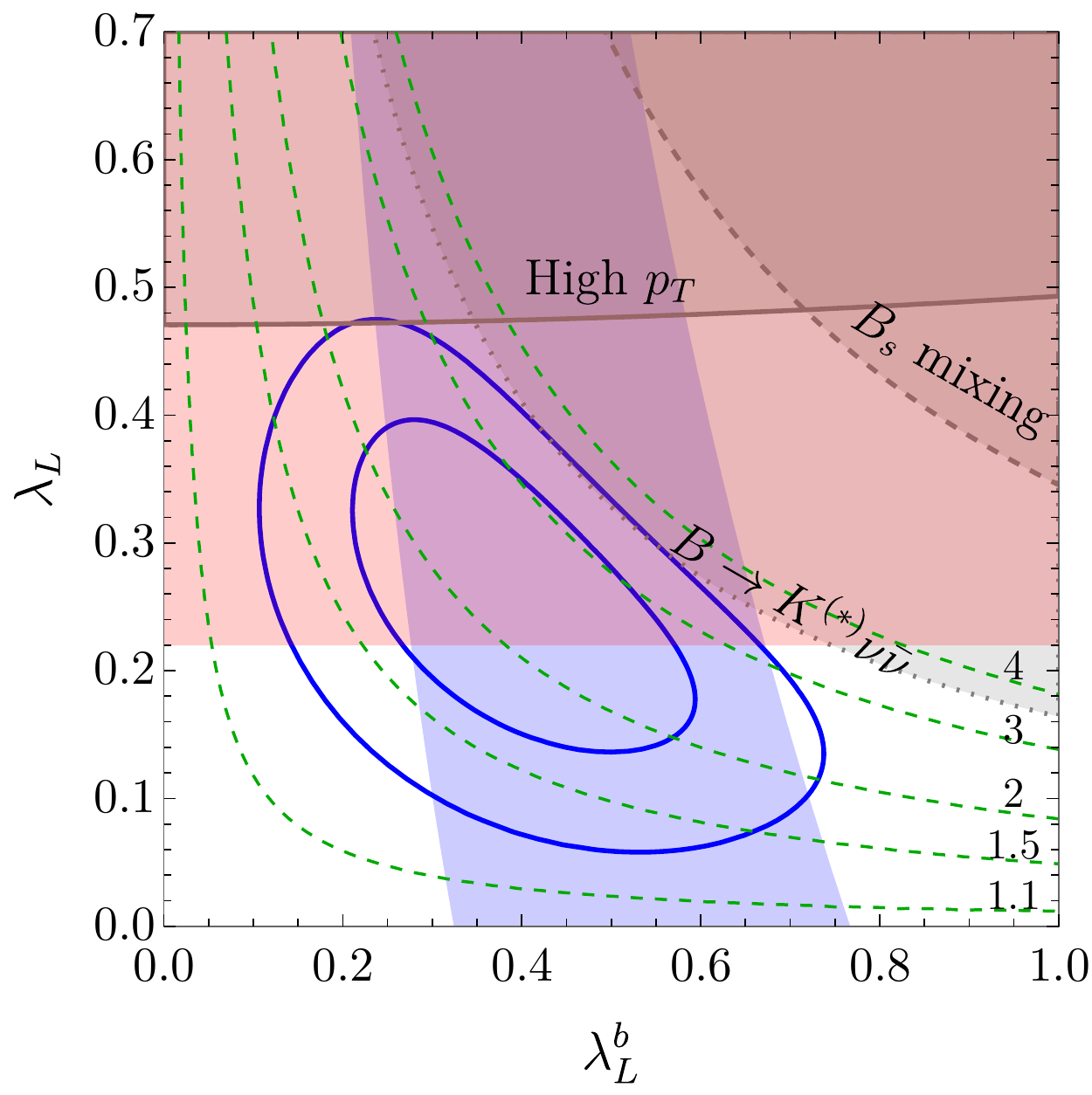}
    \includegraphics[height=0.49\textwidth]{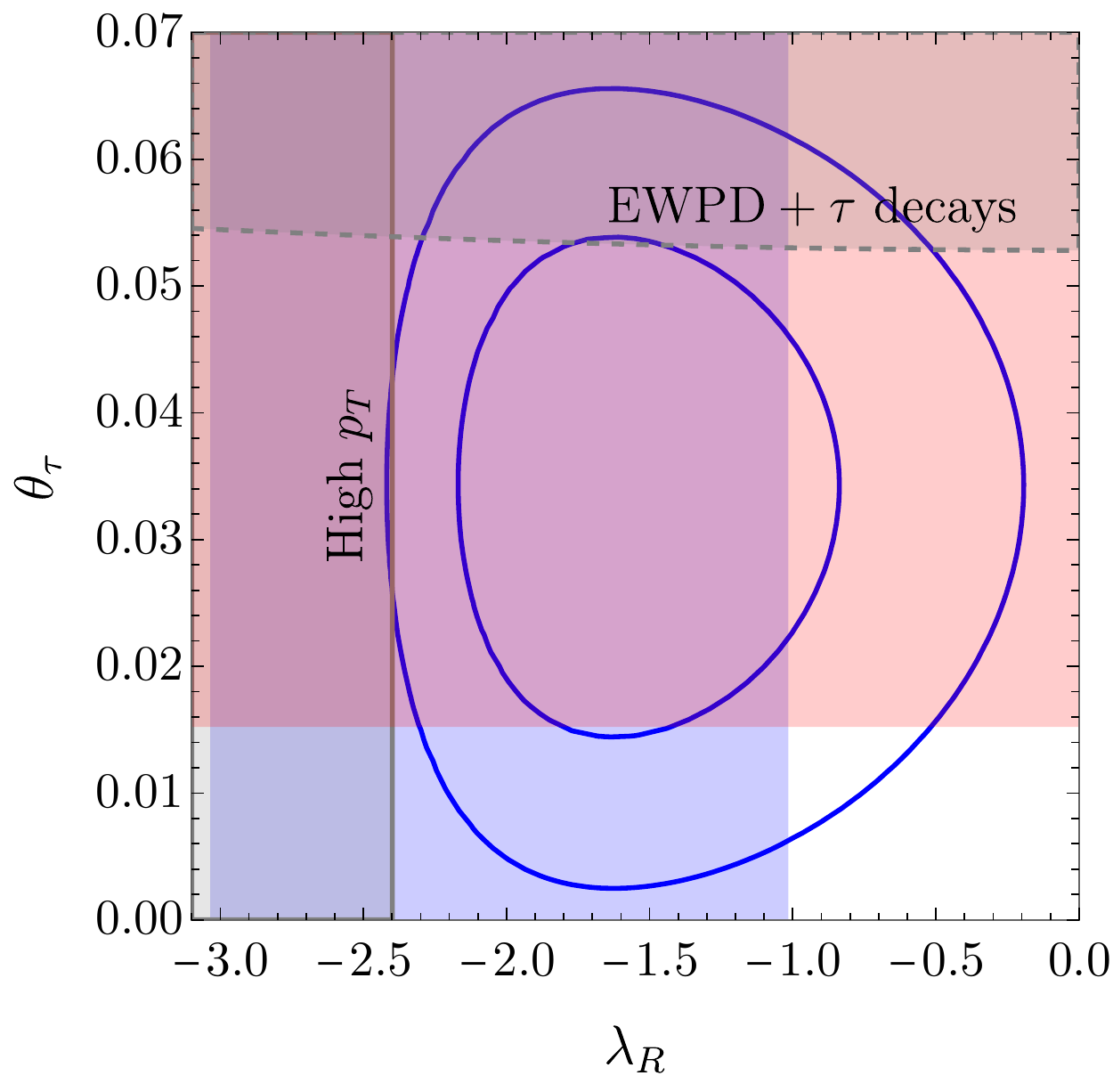}
    \caption{Global fit (solid blue lines) as a function of the couplings. We fix $M_{1} = 2$\,TeV, $\lambda_R^b=-2$, and the couplings not shown to the best-fit point.  The blue and red regions are the 1$\sigma$ preferred regions for $R_{D^{(*)},\Lambda_c}$ and the $L$-observables, respectively. The gray contours give 95\% CL exclusion limits from: $pp\rightarrow \tau\tau/\tau\nu$ (solid), $B_s$-mixing (dashed), $B\rightarrow K \nu\bar\nu$ (dotted) and EWPD together with LFU tests in $\tau$ decays (dashed). On the left panel, the green dashed lines show contours of $R_{K^{(*)}}^{\nu}$ with values indicated in the lower right corner.}
    \label{fig:globalFit}
\end{figure}

The results of our fit are shown in~\cref{fig:globalFit}. 
We can see that the most relevant constraints for the fit are $B\to K \nu \bar\nu$, $\Delta m_{B_s}$, $pp\to \tau \tau (\tau \nu)$, and EWPD with LFU tests in $\tau$-decays. The $95\%$ CL exclusion limits for the other bounds do not show because they are sub-leading constraints appearing only for larger values of the couplings.
We see that while parameter space exists where the non-leptonic puzzle and the charged-current $B$-anomalies can be simultaneously explained, their combination necessarily leads to a large effect in $B\rightarrow K \nu \bar\nu$. This occurs since the $L$-observables scale as $\theta_\tau \lambda_L$ while effects in $b\rightarrow c\tau\nu$ behave as $\lambda_L^b \lambda_R$, where $|\theta_\tau|$ and $|\lambda_R|$ are bounded by EWPO and high-$p_T$, respectively. This means one needs $\lambda_L \approx \lambda_L^b \approx  0.4$ in order to explain both puzzles, leading to a sizeable contribution to $B\rightarrow K\nu\bar\nu$ which scales as $\lambda_L \lambda_L^b$. Additionally, $b\rightarrow c\tau \nu$ transitions receive a sub-dominant vector contribution from $\lambda_L \lambda_L^b$, preferring the product to be positive. This fixes the sign of the contribution to $B\rightarrow K \nu\bar\nu$ to be positive as well, meaning we interfere constructively with the SM and go in the direction of the stronger bound given in~\cref{eq:limitBKnunu}.\footnote{A solution with negative $\lambda_L \lambda_L^b$ is also possible, which gives smaller contributions to $B\rightarrow K \nu\bar\nu$ but also a worse quality of fit for $b\rightarrow c\tau \nu$. As a result,  the overall global fit is slightly worse in this scenario.}
In particular, our best fit point predicts $R_{K^{(*)}}^{\nu}=2.3 \pm 0.5$.

While the $B\rightarrow K \nu\bar\nu$ bound allows us to do the central value called for by the charged-current $B$-anomalies if we only solve the non-leptonic puzzle at $\approx 1\sigma$, the reverse is only partially possible due to high-$p_T$ constraints requiring $|\lambda_L| \lesssim 0.5$. The strength of the high-$p_T$ bound is due to the fact that promoting $S_1$ to a doublet of $U(2)_q$ requires that one leptoquark couples to valence quarks, as shown in~\cref{fig:highpTdiag}. In summary, the fit shows that a combined explanation of both puzzles is possible, but comes with the prediction of a significantly enhanced $B\rightarrow K \nu \bar\nu$ over the SM, well within the reach of Belle II.

\section{Conclusions}
\label{sec:conc}

In this work we present a possible solution to the recently observed tensions in the optimized non-leptonic observables ($L_{K^*\bar{K}^*}$ and $L_{K\bar{K}}$) via $S_1$ scalar leptoquarks. As discussed in \cref{sec:obsNPEFT}, a solution to both non-leptonic anomalies requires a NP contribution to the Wilson coefficients of either the QCD-penguin ${\cal O}_{4s(d)}$ or the chromomagnetic dipole operators ${\cal O}_{8gs(d)}$. Here, we have presented a solution contributing to the latter given the large fine-tuning required for the former. A challenge for any model based on generating the chromomagnetic dipoles is to avoid the constraints on the corresponding electromagnetic dipoles $C_{7\gamma s(d)}$ contributing to $B \to X_{s,d} \gamma$ transitions. It is important to emphasize that this is achieved without any tuning in our model- it happens simply because the $S_1$ hypercharge has the right value to yield an accidental cancellation in the RGE mixing of the two dipoles.

The tensions in $L_{K^*\bar{K}^*}$ and $L_{K\bar{K}}$ are calling for a NP chromomagnetic dipole contribution which is of the same order as the SM one when NP contributions of opposite signs in both $b\to s$ and $b\to d$ transitions are allowed. To achieve this with TeV scale NP, one requires a significant chiral enhancement compared to the SM $m_b$ factor. In our case, this is achieved via an  $O(1)$ Yukawa coupling with a RH neutrino. Furthermore, the contribution to the chromomagnetic dipole is maximized when the RH neutrino mass is similar to that of the $S_1$ LQ, hinting at a possible connection to TeV-scale neutrino mass generation. In particular, we have shown that a multi-scale version of the inverse-seesaw mechanism for neutrino masses fits well within our setup, predicting sizable violations of PMNS unitarity in the third family, as first pointed out in~\cite{Greljo:2018tuh}.

To explain the $L$-observables, the model requires sizeable couplings of the $S_1$ LQ to light-family left-handed quarks that could imply a large breaking of the $U(2)_q$ flavor symmetry. However, by promoting the $S_1$ LQ to a doublet of $U(2)_q$, these couplings can preserve the $U(2)_q$ symmetry. Similar ideas have appeared previously in the literature in different contexts~\cite{Barbieri:2012uh, Cornella:2019hct, Greljo:2022jac}, showing that promoting NP fields to $U(2)$ multiplets can be an effective way to suppress FCNC in the light sector while allowing for sizeable contributions in third-to-second family processes. This is a common characteristic of models with minimally-broken $U(2)$ flavor symmetries~\cite{Davighi:2023evx}, and we find that the idea can be nicely implemented within our model.

Moreover, it is well-known that the $S_1$ LQ is one of the few mediators that can provide a NP explanation for hints of LFUV in charged-current $B$-meson decays, measured by the ratios $R_D$ and $R_{D^*}$. We show that an explanation of the non-leptonic puzzle can be consistently combined with an explanation of the charged-current $B$-anomalies by performing a global fit including the $L$-observables, $R_{D^{(*)}}$, and all relevant constraints such as FCNC processes, EWPO, LFU tests in $\tau$-decays, and high-$p_T$ to determine the allowed parameter space of the model. The main outcome of this fit, shown in \cref{fig:globalFit}, is that a consistent combined explanation of the non-leptonic puzzle and the charged-current $B$-anomalies predicts a large enhancement to $\mathcal{B}(B\rightarrow K \nu \bar{\nu}) $ as compared with the SM prediction. 

In summary, it is very interesting that two open puzzles in $B$-physics can be connected by the $S_1$, while simultaneously satisfying all relevant low- and high-energy constraints. On the phenomenological side, the smoking gun for the model would be a measurement of $\mathcal{B}(B\rightarrow K \nu \bar{\nu}) $ well above the SM value. This prediction is even now being tested at Belle II, which ultimately aims to measure the SM value of $\mathcal{B}(B\rightarrow K \nu \bar{\nu}) $ with a 10\% relative error.

\section*{Acknowledgements}
We thank M. Misiak for useful discussions on bounds on the electromagnetic and chromomagnetic operators and also A. Crivellin and G. Isidori for the useful discussions on the model.
J.M. gratefully acknowledges the financial support from ICREA under the ICREA Academia programme and from the Pauli Center (Zurich) and the Physics Department of University of Zurich. J.M.  also received financial support from Spanish Ministry of Science, Innovation and Universities (project PID2020-112965GB-I00) and from the Research Grant Agency of the Government of Catalonia (project SGR 1069). 
The work of J.M.L. and B.A.S. has been suported by the European Research Council (ERC) under the European Union’s Horizon 2020 research and innovation programme under grant agreement 833280 (FLAY), and by the Swiss National Science Foundation (SNF) under contract 200020-204428.

\appendix

\section{Rotation matrices}
\label{app:Rotation}

The expression for the Yukawas in the interaction basis is given in~\cref{eq:YukSpurions}. Assuming we are working in the down basis for the left-handed light families, and in the mass basis for the right-handed light families, the rotation matrices to go from the interaction basis to the mass basis are implemented as follows
\begin{equation}
Y_{u}=L^{\dagger}_{u} \,\hat Y_{u}\,R_{u},~~~~~Y_{d}=L^{\dagger}_{d} \,\hat Y_{d} \,R_{d},
\end{equation}
where $\hat Y_{u,d}$ are diagonal matrices, and the rotation matrices read
\begin{align}
L_u &= 
\begin{pmatrix}
V_{ud} && V_{us} && x_t V_{ub} \\
V_{cd} && V_{cs} && x_t V_{cb} \\
x_t V_{td} && x_t V_{ts} && 1 \\
\end{pmatrix} + O(V_{us}^4),~~~~~~~
L_d= 
\begin{pmatrix}
1 && 0 && -x_b V_{td}^* \\
0 && 1 && -x_b V_{ts}^* \\
x_b V_{td} && x_b V_{ts} && 1 \\
\end{pmatrix} + O(V_{us}^4).\\
R_u &\approx 
\begin{pmatrix}
1 && 0 && x_t\frac{m_u}{m_t} V_{ub}  \\
0 && 1 &&  x_t\frac{m_c}{m_t}V_{cb}  \\
-x_t\frac{m_u}{m_t} V^*_{ub}  && -x_t\frac{m_c}{m_t} V^*_{cb}  && 1 \\
\end{pmatrix},~
R_d\approx  
\begin{pmatrix}
1 && 0 && -x_b\frac{m_d}{m_b} V_{td}^* \\
0 && 1 && -x_b\frac{m_s}{m_b} V_{ts}^* \\
x_b\frac{m_d}{m_b} V_{td} && x_b \frac{m_s}{m_b}V_{ts} && 1 \\
\end{pmatrix}.
\end{align}

\section{Wilson coefficients for the electroweak fit}
\label{app:EWWC}

Running from the high scale to the EW scale, many of the SMEFT operators relevant for the EW fit receive contributions. Here we provide the formulas for all non-vanishing WCs involved in the EW fit and LFU tests of $\tau$ decays as function of the fundamental parameters of the model at leading-log order in $y_t$ and the gauge couplings~\cite{Jenkins:2013wua,Alonso:2013hga,Celis:2017hod}. 
For the SM parameters, we take their quadratic average along the flow: $y_t=0.88$, $g_L=0.64$ and $g_Y=0.36$ (see Appendix B of~\cite{Allwicher:2023aql} for details), performed with \texttt{DsixTools}~\cite{Celis:2017hod}. For the EW scale we take $\mu_{\rm EW} =m_t$.
Notice that $i=1,2$ and $\alpha=1,2,3$. Repeated indices in the WCs indicate diagonal elements. The relevant WCs read
\begin{align}
[C^{(1)}_{H l}]_{ii}=&\frac{g_Y^2 |y_N|^2\log \frac{\mu_{\rm EW}^2}{M_R^2}}{192 \pi^2 M_R^2},\\
[C^{(1)}_{H l}]_{33}=&
\frac{|y_N|^2}{4 M_R^2}
+\frac{3y_t^2 |y_N|^2  \log \frac{\mu_{\rm EW} ^2}{M_R^2}}{64 \pi ^2 M_R^2}+\frac{3 y_t^2 V_L^{\dagger}V_L \log \frac{\mu_{\rm EW} ^2}{M_1^2}}{64 \pi ^2
   M_1^2}\nonumber\\
   & +\frac{g_Y^2 V_L^{\dagger}V_L \log \frac{\mu_{\rm EW} ^2}{M_1^2}}{192 \pi ^2 M_1^2}+\frac{g_Y^2 |\lambda_L|^2 \log \frac{\mu_{\rm EW} ^2}{M_1^2}}{96 \pi ^2 M_1^2}+\frac{g_Y^2 |y_N|^2 \log \frac{\mu_{\rm EW} ^2}{M_R^2}}{128 \pi ^2
   M_R^2},
\end{align}
\begin{align}
[C^{(3)}_{H l}]_{ii}=&-\frac{g_L^2 |y_N|^2\log \frac{\mu_{\rm EW}^2}{M_R^2}}{192 \pi^2 M_R^2},\\
[C^{(3)}_{H l}]_{33}=&
-\frac{|y_N|^2}{4 M_R^2}
   -\frac{3 y_t^2 |y_N|^2  \log \frac{\mu_{\rm EW} ^2}{M_R^2}}{64 \pi ^2 M_R^2}
   +\frac{3 y_t^2 V_L^{\dagger}V_L \log \frac{\mu_{\rm EW} ^2}{M_1^2}}{64 \pi ^2
   M_1^2}
   \nonumber\\
   &-\frac{g_L^2 V_L^{\dagger}V_L \log \frac{\mu_{\rm EW} ^2}{M_1^2}}{64 \pi ^2 M_1^2}-\frac{g_L^2 |\lambda_L|^2 \log \frac{\mu_{\rm EW} ^2}{M_1^2}}{32 \pi ^2 M_1^2}+\frac{5 g_L^2 |y_N|^2 \log \frac{\mu_{\rm EW} ^2}{M_R^2}}{128 \pi ^2
   M_R^2},
\end{align}
\begin{align}
[C_{H e}]_{ii}=&\frac{g_Y^2 |y_N|^2\log \frac{\mu_{\rm EW}^2}{M_R^2}}{96 \pi^2 M_R^2},\\
[C_{H e}]_{33}=&
\frac{g_Y^2 |y_N|^2 \log \frac{\mu_{\rm EW} ^2}{M_R^2}}{96 \pi ^2 M_R^2}+\frac{g_Y^2 |\lambda _R|^2 \log \frac{\mu_{\rm EW} ^2}{M_1^2}}{48 \pi ^2 M_1^2},
\end{align}
\begin{align}
[C_{H q}^{(1)}]_{ii}=&-\frac{g_Y^2 |\lambda_L|^2 \log \frac{\mu_{\rm EW} ^2}{M_1^2}}{192 \pi ^2 M_1^2}-\frac{g_Y^2 |y_N|^2 \log \frac{\mu_{\rm EW} ^2}{M_R^2}}{576 \pi ^2 M_R^2},\\
[C_{H q}^{(1)}]_{i3}=&-\frac{g_Y^2V_i \lambda_L^*  \log \frac{\mu_{\rm EW} ^2}{M_1^2}}{192 \pi ^2 M_1^2},\\
[C_{H q}^{(1)}]_{33}=&-\frac{g_Y^2 V_L^{\dagger}V_L \log \frac{\mu_{\rm EW} ^2}{M_1^2}}{192 \pi ^2 M_1^2}-\frac{g_Y^2 |y_N|^2 \log \frac{\mu_{\rm EW} ^2}{M_R^2}}{576 \pi ^2 M_R^2},
\end{align}
\begin{align}
[C_{H q}^{(3)}]_{ii}=&-\frac{g_L^2 |\lambda_L|^2 \log \frac{\mu_{\rm EW} ^2}{M_1^2}}{192 \pi ^2 M_1^2}-\frac{g_L^2 |y_N|^2 \log \frac{\mu_{\rm EW} ^2}{M_R^2}}{192 \pi ^2 M_R^2},\\
[C_{H q}^{(3)}]_{i3}=&-\frac{g_L^2V_i \lambda_L^*  \log \frac{\mu_{\rm EW} ^2}{M_1^2}}{192 \pi ^2 M_1^2},\\
[C_{H q}^{(3)}]_{33}=&-\frac{g_L^2 V_L^{\dagger}V_L \log \frac{\mu_{\rm EW} ^2}{M_1^2}}{192 \pi ^2 M_1^2}-\frac{g_L^2 |y_N|^2 \log \frac{\mu_{\rm EW} ^2}{M_R^2}}{192 \pi ^2 M_R^2},
\end{align}
\begin{align}
[C_{H u}]_{\alpha \alpha}=& -\frac{g_Y^2 |y_N|^2 \log \frac{\mu_{\rm EW} ^2}{M_R^2}}{144 \pi ^2 M_R^2},\\
[C_{H d}]_{\alpha \alpha}=&\frac{g_Y^2 |y_N|^2 \log \frac{\mu_{\rm EW} ^2}{M_R^2}}{288 \pi ^2 M_R^2}.
\end{align}

\section{Charged-current transitions}
\label{app:ChargeCurrent}

Following~\cite{Gherardi:2020qhc,Iguro:2018vqb}, contributions to $B_{c,u}\to \tau \nu$ in our model read
\begin{align}
\frac{{\rm Br}(B_c\to \tau \nu)}{{\rm Br}(B_c\to \tau \nu)_{\rm SM}}=
\left|
1+C^c_{V_L}-\chi_{B_c}  C^c_{S_L}
\right|^2,\\
\frac{{\rm Br}(B_u\to \tau \nu)}{{\rm Br}(B_u\to \tau \nu)_{\rm SM}}=
\left|
1+C_{V_L}^u-\chi_{B_u}  C_{S_L}^u
\right|^2,
\end{align}
where $\chi_{B_c}=m_{B_c}^2/[m_{\tau}(m_b+m_u)]\approx 4.3$, 
$\chi_{B_u}=m_{B_u}^2/[m_{\tau}(m_b+m_u)]\approx 3.8$, $C^c_{V_L}$ and $C^c_{S_L}$ are given in~\cref{eq:CVc,eq:CSc}, and
\begin{align}
C_{V_L}^u (m_b) &= - \frac{v^2}{V_{ub}} \left(\sum_{i=1}^{2} V_{1i}[C_{lq}^{(3)}]_{33i3}
+x_t V_{ub} [C_{lq}^{(3)}]_{3333}
\right)\approx 
 \frac{v^2}{V_{ub}}
\frac{\lambda_L^*\lambda_L^{b}}{4M_1^2}(V_{ud}\epsilon_L+ V_{us} )
,\\
C_S^u (m_b)&= -\eta_S \frac{v^2}{2V_{ub}}[C_{lequ}^{(1)}]_{3331}^*=
 -\eta_S \frac{v^2}{V_{ub}}\frac{m_u}{m_c} \frac{\lambda^*_R \lambda_L^b}{4M_1^2}(V_{ud}\epsilon_L + V_{cd})\approx 0,
\end{align}
with $\eta_S \approx 1.7$.
The current experimental measures and SM predictions are~\cite{Akeroyd:2017mhr,ParticleDataGroup:2022pth,UTfit:2022hsi, Gherardi:2020qhc}
\begin{align}
{\rm Br}(B_c\to \tau \nu)_{\rm Exp}=&\,2.3\%,~~~~~~~~~~~~~~
{\rm Br}(B_c\to \tau \nu)_{\rm SM}<\, 10\%, ~~~(95\%~{\rm C.L.}),\\
{\rm Br}(B_u\to \tau \nu)_{\rm Exp}=&1.09(24)\cdot 10^{-4},~~
{\rm Br}(B_u\to \tau \nu)_{\rm SM}=\,0.869(47)\cdot 10^{-4}.
\end{align}
Likewise, for $D^+_{(s)}\to \tau \nu$, we get
\begin{align}
\frac{{\rm Br}(D_s\to \tau \nu)}{{\rm Br}(D_s\to \tau \nu)_{\rm SM}}=
\left|
1+C_{V_L}^{cs}-\chi_{D_s} C_{S_L}^{cs}
\right|^2,\\
\frac{{\rm Br}(D^+\to \tau \nu)}{{\rm Br}(D^+\to \tau \nu)_{\rm SM}}=
\left|
1+C_{V_L}^{cd}-\chi_{D}  C_{S_L}^{cd}
\right|^2,
\end{align}
where $\chi_{D_s}=m_{D_s}^2/[m_{\tau}(m_s+m_c)]\approx 1.6$, 
$\chi_{D}=m_{D}^2/[m_{\tau}(m_d+m_c)]\approx 1.5$, and
\begin{align}
C_{V_L}^{cs} (m_c) &= - \frac{v^2}{V_{cs}} \left(\sum_{i=1}^{2} V_{2i}[C_{lq}^{(3)}]_{33i2}
+x_t V_{cb} [C_{lq}^{(3)}]_{3332}
\right)\approx 
 v^2
\frac{|\lambda_L|^2}{4M_1^2}
,\\
C_S^{cs} (m_c)&= -\eta^{\prime}_S \frac{v^2}{2V_{cs}}[C_{lequ}^{(1)}]_{3322}^*=
-\eta^{\prime}_S v^2 \frac{\lambda_L \lambda_R^*}{4M_1^2},
\\
C_{V_L}^{cd} (m_c) &= - \frac{v^2}{V_{cd}} \left(\sum_{i=1}^{2} V_{2i}[C_{lq}^{(3)}]_{33i1}
+x_t V_{cb} [C_{lq}^{(3)}]_{3331}
\right)\approx 
 v^2
\frac{|\lambda_L|^2}{4M_1^2}
,\\
C_S^{cd} (m_c)&= -\eta^{\prime}_S \frac{v^2}{2V_{cd}}[C_{lequ}^{(1)}]_{3312}^*=
-\eta^{\prime}_S v^2 \frac{\lambda_L \lambda_R^*}{4M_1^2},
\end{align}
where $\eta^{\prime}_S\approx 2$~\cite{Gherardi:2020qhc}.
The current experimental measurements~\cite{ParticleDataGroup:2022pth} and SM predictions, taken from \texttt{flavio}~\cite{Straub:2018kue}, are 
\begin{align}
{\rm Br}(D_s\to \tau \nu)_{\rm Exp}=&\,5.48(23)\%,~~~~~~~~~~~~~~~~~~
{\rm Br}(D_s\to \tau \nu)_{\rm SM}=\,5.32(4)\%,\\
{\rm Br}(D^+\to \tau \nu)_{\rm Exp}<&\,1.2\cdot 10^{-3},~(90\%~{\rm C.L.}),~~
{\rm Br}(D^+\to \tau \nu)_{\rm SM}=\,1.09(1)\cdot 10^{-3}.
\end{align}

\bigskip

\bibliographystyle{JHEP}
\bibliography{note}

\end{document}